\documentclass[twocolumn,prx,superscriptaddress]{revtex4}
\usepackage{amsfonts}
\usepackage{amsmath}
\usepackage{amssymb}
\usepackage{graphicx}
\usepackage{bm}
\usepackage{color}
\usepackage{mathrsfs}
\usepackage{amsbsy}
\usepackage{latexsym,epsfig,graphicx}
\usepackage{dcolumn}
\usepackage{subfigure}
\usepackage{comment}
\usepackage{xspace}
\usepackage{epstopdf}
\usepackage{tabularx}
\usepackage{longtable}
\usepackage[colorlinks=true, letterpaper=true, pdfstartview=FitV, linkcolor=blue, citecolor=blue, urlcolor=blue]{hyperref}
\usepackage[normalem]{ulem}
\newcommand{\RNum}[1]{\uppercase\expandafter{\romannumeral #1\relax}}

\begin{document}
\title{Exact solution of non-Hermitian systems with generalized boundary conditions: size-dependent boundary effect and fragility of skin effect}

\author{Cui-Xian Guo}
\affiliation{Beijing National Laboratory for Condensed Matter Physics,
Institute of Physics, Chinese Academy of Sciences, Beijing 100190, China}

\author{Chun-Hui Liu}
       \affiliation{Beijing National Laboratory for Condensed Matter Physics, Institute of Physics, Chinese Academy of Sciences, Beijing 100190, China}
\affiliation{School of Physical Sciences, University of Chinese Academy of Sciences, Beijing 100049, China}

\author{Xiao-Ming Zhao}
\affiliation{Beijing National Laboratory for Condensed Matter Physics,
Institute of Physics, Chinese Academy of Sciences, Beijing 100190, China}

\author{Yanxia Liu}
\affiliation{Beijing National Laboratory for Condensed Matter Physics,
Institute of Physics, Chinese Academy of Sciences, Beijing 100190, China}

\author{Shu Chen}
\email{schen@iphy.ac.cn}
\affiliation{Beijing National Laboratory for Condensed Matter Physics,
Institute of Physics, Chinese Academy of Sciences, Beijing 100190, China}
\affiliation{School of Physical Sciences, University of Chinese Academy of Sciences, Beijing 100049, China}
\affiliation{Yangtze River Delta Physics Research Center, Liyang, Jiangsu 213300, China}
\begin{abstract}

Systems with non-Hermitian skin effects are very sensitive to the imposed boundary conditions and lattice size, and thus an important question is whether non-Hermitian skin effects can survive when deviating from the open boundary condition. To unveil the origin of boundary sensitivity,
we present exact solutions for one-dimensional non-Hermitian models with generalized boundary conditions and study rigorously the interplay effect of lattice size and boundary terms. Besides the open boundary condition, we identify the existence of non-Hermitian skin effect when one of the boundary hopping terms vanishes. Apart from this critical line on the boundary parameter space, we find that the skin effect is fragile under any tiny boundary perturbation in the thermodynamic limit, although it can survive in
a finite size system. Moreover, we demonstrate that the non-Hermitian Su-Schreieffer-Heeger model exhibits a new phase diagram in the boundary critical line, which is different from either open or periodical boundary case.

\end{abstract}
\maketitle

{\it Introduction.-}
It is well known that the spectrum of a periodic crystal can be characterized by the Bloch wave
vector and the periodic boundary condition (PBC) is usually taken for the convenience of calculating the band structure \cite{Ashcroft}. If the system size is large enough, the bulk spectrum is stable against boundary perturbations even though the translation invariance of the system is broken \cite{Alase,Alase2,Kunst-PRB1029}. This constitutes the foundation for understanding why the bulk energy levels of a large system with open boundary condition (OBC)   can be reproduced from the Bloch band calculation. However, such a paradigm is challenged in some non-Hermitian systems \cite{Alvarez,TELee,Xiong,SYao1,Leykam}, for which the wave functions in large systems with OBC  accumulate on the boundary accompanying with a remarkably different eigenvalue spectrum
from the periodic system \cite{SYao2,Kunst,KZhang,KYokomizo,LeeCH,Okuma,HShen}. This phenomenon is coined as the non-Hermitian skin effect (NHSE) \cite{SYao1}  and recently attracted intensive studies \cite{Slager,HJiang,WYi,LJin,Kou,Sato,Zhou,CHLiu2,Gong,Longhi-PRR,Herviou,ZSYang,GongJB,YXLiu,Ezawa,YFYi,Imura}.

The NHSE suggests that the change of boundary condition may induce dramatic change of bulk properties of non-Hermitian systems \cite{SYao1,Turker,Ueda,SYao2,Kunst,KZhang,KYokomizo,LeeCH,Okuma,HShen,Budich,Sato-PRB,Kunst-PRB,RChen,Budich-EPJD}.
Size-dependent NHSEs are also observed in some coupled non-Hermitian chains \cite{CHLiu2020,CSE} and non-reciprocal chains with impurity \cite{Linhu,Longhi-Adp}. These studies indicate that both boundaries and lattice size play an important role in these boundary sensitive effects. Although the spectral flow from PBC to OBC is studied by introducing an imaginary flux \cite{LeeCH,LeeCH-PRB2020}, it is still elusive to get a quantitative understanding of the sharp change of spectrum and wave functions of skin modes under tiny boundary perturbations. A  more challenging task is to count quantitatively the interplay effect of system size and boundary perturbations and  unveil the intrinsic reason behind the boundary sensitive effects.
As numerical methods for boundary sensitive problems are time consuming and sometimes unreliable due to the
existence numerical errors and calculation precision \cite{Colbrook,Reichel}, exact solutions are highly desirable for analytically exploring the size-dependent boundary effect.

In this letter, we present exact solutions of non-Hermitian models with non-reciprocal hopping under generalized boundary conditions (GBCs), which enable us to explore rigorously the interplay effect of lattice size and boundary perturbations.
Our analytical results show explicitly how the lattice size and boundary terms affect the solutions of eigen equations.
Particularly, we find the existence of NHSE in a critical line on the boundary parameter space, including the OBC as a special case. Apart from the critical line, the NHSE is unstable against any tiny boundary perturbations in the thermodynamic limit and thus is fragile, although it may survive in a finite size system. Moreover, we find that the two-band system can exhibit a new phase diagram in the critical line, which is different from either PBC or OBC case, but is a combination of the two cases. Our work demonstrates  novel phenomena induced by the boundary terms from the perspective of exact solution and provides a firm ground for understanding boundary sensitivity phenomena in non-Hermitian systems.

\begin{figure*}[tbp]
\includegraphics[width=1.0\textwidth]{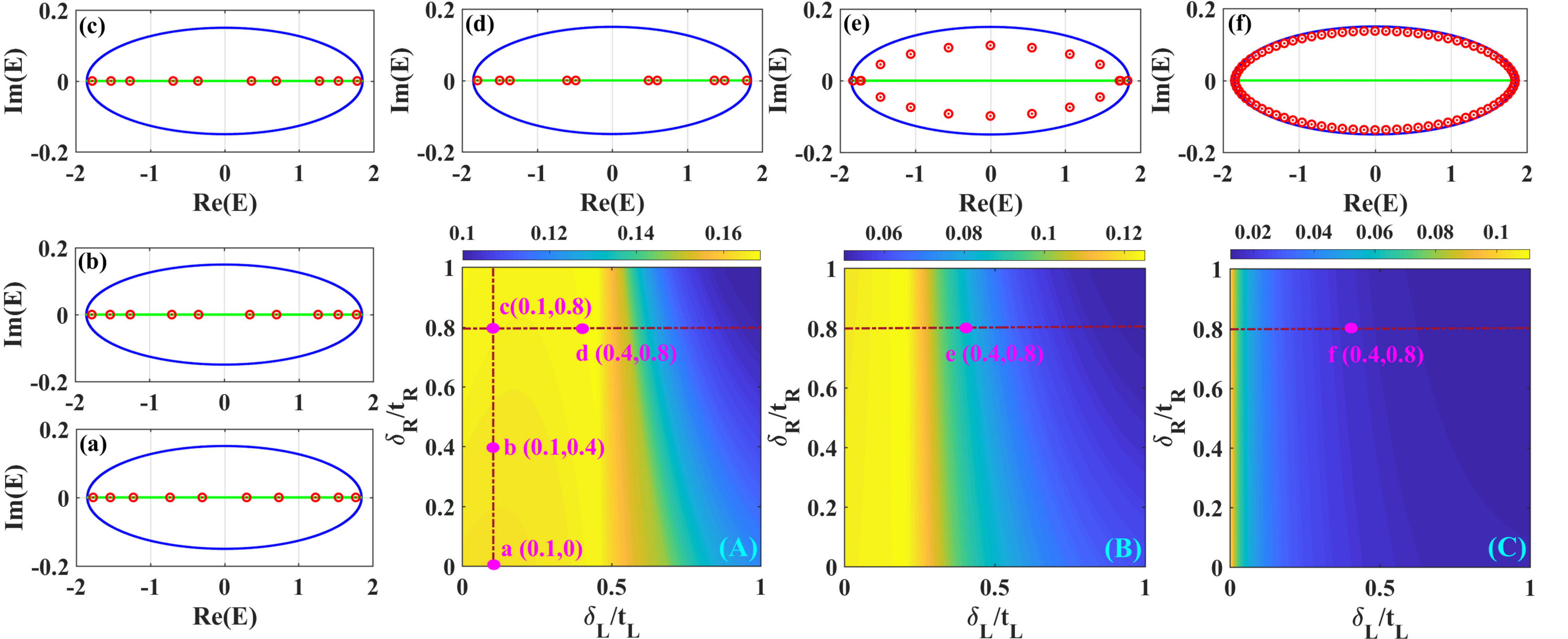}
\caption{(A-C) $\overline{\mathrm{IPR}}$ on the parameter space of $\delta_L/t_L$ and $\delta_R/t_R$ for HN model with $N=10,20,80$, respectively. (a-f) Energy spectrum (red circles and dots) corresponding dots 'a-f' in (A-C), respectively. The analytical results (red circles) are in exact agreement with the numerical results (red dots). The green and blue line represents energy spectrum corresponding to  OBC and PBC case in the thermodynamic limit, respectively. Common parameters: $t_{L}=1, t_{R}=0.85$.}%
\label{fig1}
\end{figure*}

{\it Hatano-Nelson model with generalized boundary conditions.-} We start with the Hatano-Nelson (HN) model \cite{Hatano,HatanoPRB} with GBC described by
\begin{equation}
\hat{H}=\sum\limits_{n=1}^{N-1}\left[ t_{L}\hat{c}_{n}^{\dag
}\hat{c}_{n+1}+t_{R}\hat{c}_{n+1}^{\dag }\hat{c}_{n}\right] +\delta _{R}\hat{c}_{1}^{\dag
}\hat{c}_{N}+\delta _{L}\hat{c}_{N}^{\dag }\hat{c}_{1},
\end{equation}
where $N$ is the number of lattice sites, $\delta_{L}, \delta_{R}\in \mathbb{R}$ determines the GBCs, and $t_{L}, t_{R}\in \mathbb{R}$ are imbalanced hopping amplitudes which can
be parameterized as $t_L = te^{-g}$ and $t_R = te^g$ with real $t$ and $g$.
This is the minimal model which can display nontrivial size-dependent boundary effect.

The corresponding eigenvalue equation can be written as
$
\hat{H}|\Psi\rangle = E|\Psi\rangle
$,
where $|\Psi\rangle= \sum_{n} \psi_n |n\rangle$ with $|n\rangle=\hat{c}_n^{\dag}|0\rangle ~(n=1,\cdots,N)$. The above eigenvalue equation consists of a series of equations, including bulk equations as follows
\begin{equation}\label{bk}
t_{R}\psi _{s}-E\psi _{s+1}+t_{L}\psi _{s+2}=0
\end{equation}%
with $s=1,2,\cdots,N-2$, and the boundary equations given by
$-E\psi _{1}+t_{L}\psi _{2}+\delta _{R}\psi _{N} =  0$ and
$\delta _{L}\psi _{1}+t_{R}\psi _{N-1}-E\psi _{N} =  0$.
By comparing the above two equations with Eq.(\ref{bk}), they are equivalent to the following boundary conditions
\begin{equation}
t_{R}\psi _{0} =\delta _{R}\psi _{N}, ~~~~~
\delta _{L}\psi _{1} = t_{L}\psi _{N+1}.  \label{bd3}
\end{equation}
Due to spatial translational property from bulk equations, we set the ansatz of wave function $\Psi _{i}$ which satisfies the bulk equations Eq.(\ref{bk}) as follows
\begin{equation}\label{Fii}
\Psi _{i}=(z_{i},z_{i}^{2},z_{i}^{3},\cdots ,z_{i}^{N-1},z_{i}^{N})^{T}.
\end{equation}
By inserting  Eq.(\ref{Fii}) into the bulk equation Eq.(\ref{bk}), we obtain the expression of eigenvalue in terms of $z_i$:
\begin{eqnarray}\label{Ez1}
E &=&\frac{t_{R}}{z_{i}}+t_{L}z_{i} .
\end{eqnarray}
For a given $E$, there are two solutions $z_i$ ($z_1, z_2$), and thus they should fulfill the following constraint condition:
\begin{equation}\label{z1z2}
z_{1}z_{2}=\frac{t_{R}}{t_{L}} .
\end{equation}
Therefore, the superposition of two linearly independent solutions is also the solution of Eq.(\ref{bk}) corresponding the same eigenvalue, i.e.,
$\Psi=c_{1}\Psi _{1}+c_{2}\Psi _{2} = (\psi _{1},\psi _{2},\cdots ,\psi _{N})^{T}~~~ \label{Wave}$,
where $\psi _{n}=\sum_{i=1}^{2}(c_{i}z_{i}^{n})=c_{1}z_{1}^{n}+c_{2}z_{2}^{n}$
with $n=1,2,\cdots ,N$.

To solve the eigen equation, the general ansatz of wave function should satisfy the boundary conditions.
By inserting  the expression of $\Psi$ into Eqs.(\ref{bd3}),  the boundary equations transforms into
$H_{B} (c_{1}, c_{2})^T =0 $ with
\begin{equation*} \label{bb1}
H_{B} =\left(
\begin{array}{cc}
t_{R}-\delta _{R}z_{1}^{N} & t_{R}-\delta _{R}z_{2}^{N} \\
z_{1}\left( \delta _{L}-t_{L}z_{1}^{N}\right)  & z_{2}\left( \delta
_{L}-t_{L}z_{2}^{N}\right)
\end{array}%
\right) .
\end{equation*}
The condition for the existence of nontrivial solutions for $(c_1, c_2)$,
including $(c_1\neq0, c_2\neq0)$ and $(c_i\neq0, c_j=0) (i,j=1,2)$, is determined by $\mathrm{det}[H_{B}] =0$,
which gives rise to the general solution: 
\begin{equation}
\begin{split}\label{qqz1}
&(z_{1}^{N+1}-z_{2}^{N+1})-\frac{\delta _{R}\delta _{L}}{t_{L}^{2}}%
(z_{1}^{N-1}-z_{2}^{N-1})\\
&-\left[ \frac{\delta _{L}}{t_{L}}+\frac{\delta _{R}%
}{t_{R}}\left( \frac{t_{R}}{t_{L}}\right) ^{N}\right] (z_{1}-z_{2})=0 .
\end{split}
\end{equation}
Eq.(\ref{qqz1}) and  Eq.(\ref{z1z2}) together determine the solution of $z_1$ and $z_2$ exactly. The solutions of $z_1$ and $z_2$ give the {\it finite-size generalized Brillouin zone} \cite{note-FGBZ,SYao1,KYokomizo,KZhang}, which may be different for different lattice size.
According to the constraint condition of Eq.(\ref{z1z2}), we can always set the solution as
\begin{equation}
z_{1}=re^{i\theta}, ~~~ z_{2}=re^{-i\theta} \label{solution-set}
\end{equation}
with $r=\sqrt{\frac{t_{R}}{t_{L}}}=e^g$, then Eq.(\ref{qqz1})  becomes
\begin{equation}\label{eq-theta}
\sin[(N+1)\theta]-\eta_1\sin[(N-1)\theta]-\eta_2\sin[\theta]=0,
\end{equation}
where $\eta_1=\frac{\delta_R\delta_L}{t_Rt_L}$ and $\eta_2=\frac{\delta _{L}}{t_{L}} r ^{-N}+\frac{\delta _{R}}{t_{R}} r ^{N}=\frac{\delta _{L}}{t_{L}} e^{-gN}+\frac{\delta _{R}}{t_{R}}
e^{gN}$.  The corresponding eigenvalue is given by
\begin{equation}
E=2\sqrt{t_{R}t_{L}}\cos \theta . \label{spectrum1}
\end{equation}
The solutions $\theta$ of Eq.(\ref{eq-theta}) may take real or complex depending on the values of $\eta_1$ and $\eta_2$. In the presence of both nonzero boundary terms, i.e., with fixed $\delta_{L,R} \neq 0$, $\eta_2$ always increases exponentially with $N$ \cite{note}, and thus the solutions are very sensitive to even a tiny boundary perturbation since the perturbation is amplified exponentially by a factor $e^{gN}$ ($g>0$) or $e^{-gN}$ ($g<0$), which is the origin of size-dependent boundary sensitivity. Such a {\it size-enhancing boundary sensitivity} has no correspondence in the Hermitian limit with $r=1$ ($g = 0$).

The OBC corresponds to the special case with $\delta _{R}=\delta _{L}=0$,
for which we have $\eta_1=\eta_2=0$ and Eq.(\ref{eq-theta}) has N real solutions given by $\theta=\frac{m\pi }{N+1}$ ($m=1,\cdots ,N$).
The corresponding eigenvalues are real with eigenstates given by
$\Psi=\left(r \sin[\theta],r^2\sin[2\theta] ,\cdots, r^N\sin[N\theta] \right)^{T}$.

For cases with either $\delta _{R}=0$ ($\delta _{L}\neq0$) or $\delta _{L}=0$ ($\delta _{R} \neq 0)$, we have $\eta_1=0$ and $\eta_2=\frac{\delta _{L}}{t_{L}} r ^{-N}$ or $\eta_2=\frac{\delta _{R}}{t_{R}} r ^{N}$.
As long as $\left|\eta_2\right|<1$, Eq.(\ref{eq-theta}) has $N$ real solutions, and the corresponding eigenvalues given by Eq.(\ref{spectrum1}) are real. Particularly, in the thermodynamic limit we have $\left|\eta_2\right|\rightarrow 0$ for the case of $\delta_R=0$ (a fixed $\delta _{L}$) and $r>1$ ($|t_R|>|t_L|$) or $\delta_L=0$ (a fixed $\delta _{R}$) and $r<1$ ($|t_R|<|t_L|$), and the solutions $\theta=\frac{m\pi}{N+1}$ are identical to the OBC case.
The analytical results indicate clearly that in these cases the system exhibits NHSE as all wavefunctions accumulate either on the left ($r<1$) or right ($r>1$) edge in the large size limit.

Now we consider the general case with nonzero $\delta_L$ and $\delta_R$. In the region of $0<\frac{\delta_R}{t_R}, \frac{\delta_L}{t_L}<1$, we have $0<\eta_1< 1$ and $\eta_2>0$.
When $\eta_2<1+\eta_1$,  Eq.(\ref{eq-theta}) has $N$ real solutions. When $\eta_2>N+1-\eta_1(N-1)$ which always holds true in the large N limit, Eq.(\ref{eq-theta}) has no real solutions but $N$ complex solutions, and the corresponding eigenvalues are complex. In this case, we have $|z_{1}|\neq|z_{2}|$.
In the thermodynamic limit, we have $|z_{1/2}| \rightarrow 1$ and $|z_{2/1}| \rightarrow \frac{t_R}{t_L}$, suggesting that the spectrum  approaches to the periodic spectrum \cite{supp}.

To give a concrete example, we display the energy spectra and averaged inverse participation ratio (IPR) in Fig.\ref{fig1} for the case of $t_{R}/t_{L}<1$ in the parameter region of $0\leq\frac{\delta _{R}}{t_R}, \frac{\delta _{L}}{t_L} \leq 1$. The averaged IPR is defined as
$
\overline{\mathrm{IPR}}=\frac{1}{N}\sum_{s=1}^{N}\mathrm{IPR_s}=\frac{1}{N}\sum_{s=1}^{N}
\frac{\sum_n|\langle n|\Psi^s \rangle|^4}{(\langle \Psi^s|\Psi^s\rangle)^2},
$
where $\Psi^s$ is the $s$-th right eigenstate $\Psi$ of $H$. 
While $\overline{\mathrm{IPR}} \sim \frac{1}{N}$ approaches zero in large N limit for homogeneously distributed eigenstates,  a finite $\overline{\mathrm{IPR}}$ gives signature of NHSE.
As shown in Fig.\ref{fig1}(A) for $N=10$, the eigenstates in the yellow region are similar to the OBC case, and the corresponding eigenvalues are real as displayed in Fig.\ref{fig1}(a)-(d). When we increase the size N, the yellow region becomes narrow. The eigenvalues with the same parameters as in Fig.\ref{fig1}(d) become complex as displayed in Fig.\ref{fig1}(e) and (f) for $N=20$ and $80$, respectively. Particularly, for $N=80$, we see that the spectrum  almost completely overlaps with the PBC spectrum,
and the blue region  almost spreads over the whole parameter space except a very narrow region near the axis of $\delta_L=0$, which is consistent with our analytic prediction.


\begin{figure}[tbp]
\includegraphics[width=0.48\textwidth]{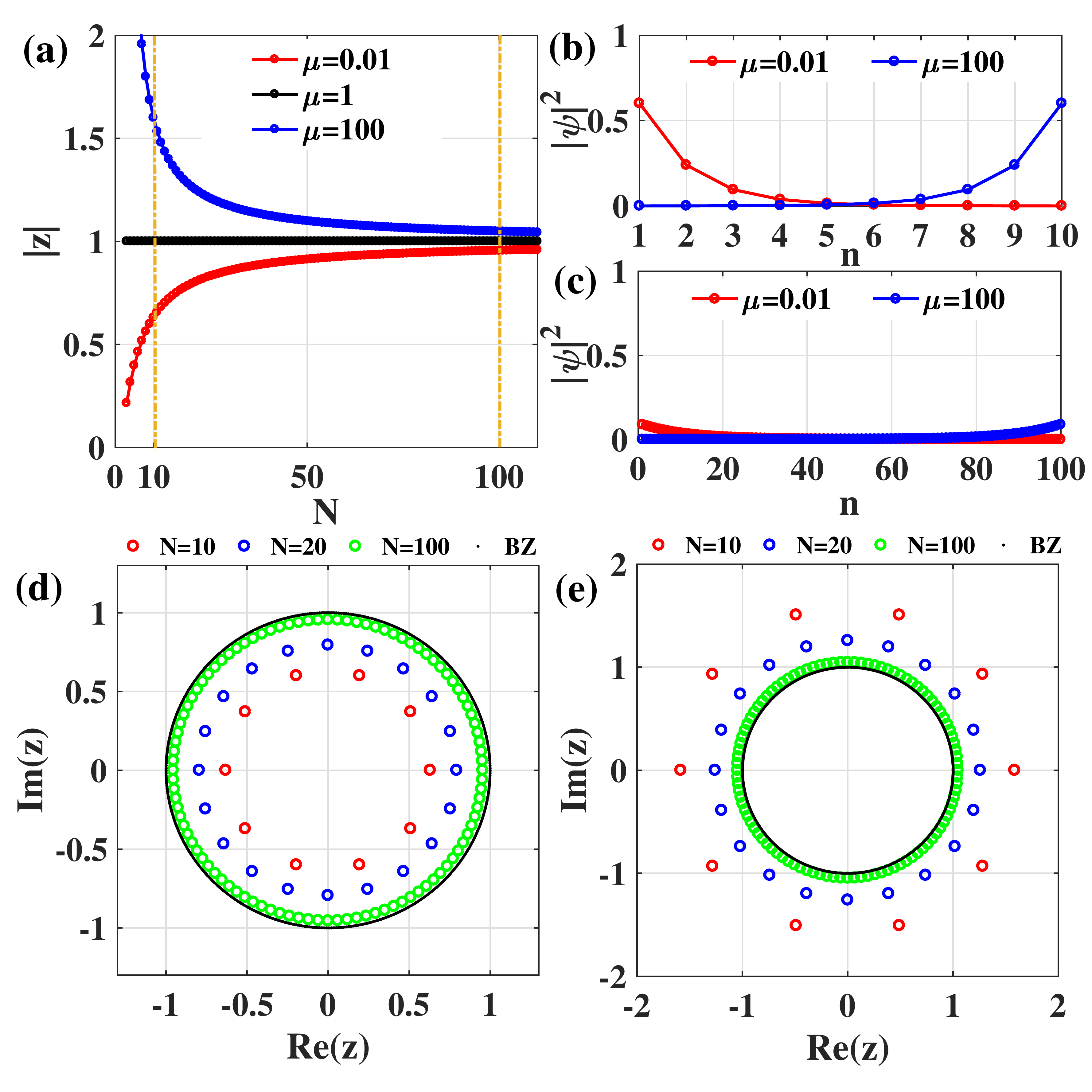}
\caption{(a) $|z|$ as a function of lattice size $N$ for HN model with $\mu=0.01,1,100$. (b, c) The profile of all eigenstates with $N=10$ and $N=100$ independent of $t_{R},t_{L}$, respectively. (d, e) The finite-size generalized Brillouin zones $z$ with $\mu=0.01$ and $\mu=100$ for different size $N=10,20,100$ independent of $t_{R},t_{L}$, respectively. The curve formed by black dots represents Brillouin zone for PBC case.}
\label{fig2}
\end{figure}

It is noticed that the solutions of $z_i$ for $c_1\neq 0$ and $c_2=0$ (or $c_1=0, c_2\neq0$) can be easily obtained by another simplified way. In this case, eigenfunction is composed of only one solution, i.e.,
$|\Psi\rangle=c_{1}|\Psi _{1}\rangle$,
and the boundary equation $H_B (c_1, 0)^T =0$ requires that
\begin{equation}
t_{R} =\delta_{R}z_1^{N}, ~~~~ \delta_{L} =t_{L}z_1^{N},
\end{equation}
which can be satisfied simultaneously only if
\begin{equation}\label{q1cdt}
\frac{t_{R}}{\delta _{R}}=\frac{\delta _{L}}{t_{L}}=\mu.
\end{equation}
We note that the special case $\mu=1$ just corresponds to the PBC.
Under the boundary condition (\ref{q1cdt}), the solution of $z_1$ is determined by
$z_1^{N}=\mu$,
which gives rise to
$z_1=\sqrt[N]{\mu }e^{i\frac{2m\pi }{N}},\text{ \ }(m=1,2,\cdots ,N)$.
It then follows that the energy spectrum is given by
\begin{equation}
E=(t_{L}\sqrt[N]{\mu }+\frac{t_{R}}{\sqrt[N]{\mu }})\cos (\theta
)+i(t_{L}\sqrt[N]{\mu }-\frac{t_{R}}{\sqrt[N]{\mu }})\sin (\theta )
\end{equation}%
with $\theta=\frac{2m\pi}{N}$,
and eigenstates as
$
\Psi= \left(\sqrt[N]{\mu }e^{i\theta},\left( \sqrt[N]{\mu }e^{i\theta} \right)^{2},\cdots,\left( \sqrt[N]{\mu }e^{i\theta} \right)^{N} \right)^{T}.
$
While we have always $|z_1|=1$ under the PBC case,  for the general case,
$
|z_1|=\sqrt[N]{\mu }
$
is not equal to 1. 
While the system may exhibit NHSE for a finite $N$, the NHSE will disappear in the large size limit as $|z_1|$ always approaches 1 when  $N \rightarrow \infty$ for a fixed $\mu$, as displayed in Fig.\ref{fig2} (here $z=z_1$). Therefore, this case is similar to the PBC case in the thermodynamic limit. If we take $\mu=r^N$, we have $E=2 \sqrt{t_L t_R} \cos \theta$ with  $\theta=\frac{2m\pi}{N}$. This special case is the so called modified PBC studied in Ref.{\cite{Imura}}. When $\mu=r^{2N}$, the spectra  $E=(t_{L}+t_{R})\cos (\theta)-i(t_{L}-t_{R})\sin (\theta )$ with $\theta=\frac{2m\pi}{N}\text{ \ }(m=1,2,\cdots ,N)$ are the same with the spectra under PBC, and we call it {\it pseudo-PBC} as the corresponding wave functions exhibit NHSE. We also present the spectra flow of the HN model in the supplementary material (SM) \cite{supp}.

\begin{figure}[tbp]
\includegraphics[width=0.48\textwidth]{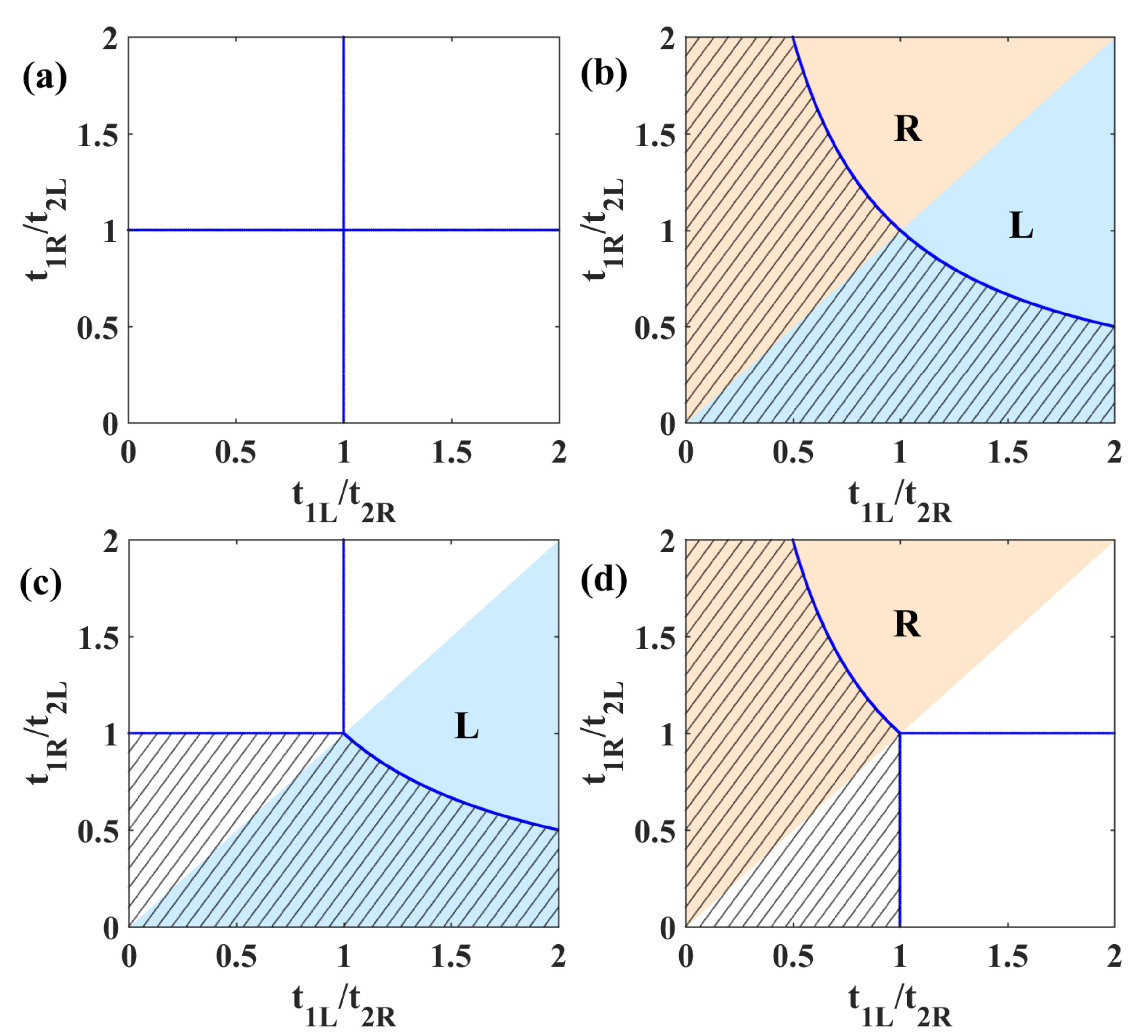}
\caption{Phase diagram for non-Hermitian SSH model (a) PBC case; (b) OBC case; (c) case of $\delta_L=0, \delta_R\neq0$; (d) case of $\delta_R=0, \delta_L\neq0$. The phase boundaries are denoted by blue lines, and  bound states exist in the shadow region. The bulk states in the blue region and orange region are located at the left and right edge, respectively. There is no NHSE in the white region.}%
\label{fig3}
\end{figure}

{\it Non-Hermitian Su-Schrieffer-Heeger model.-} We can also exactly solve the one-dimensional (1D) non-Hermitian Su-Schrieffer-Heeger (SSH) model with GBC, described by
\begin{equation}
\begin{split}
\hat{H}=&\sum\limits_{n}[ t_{1L}\hat{c}_{nA}^{\dag
}\hat{c}_{nB}+t_{1R}\hat{c}_{nB}^{\dag }\hat{c}_{nA}+t_{2R}\hat{c}_{(n+1)A}^{\dag
}\hat{c}_{nB}\\
&+t_{2L}\hat{c}_{nB}^{\dag }\hat{c}_{(n+1)A}]
+\delta _{R}\hat{c}_{1A}^{\dag
}\hat{c}_{MB}+\delta _{L}\hat{c}_{MB}^{\dag }\hat{c}_{1A} ,
\end{split} \label{GSSH}
\end{equation}
where $t_{1L/1R}$ and $t_{2L/2R}$ are imbalanced hopping term between intracell and  intercell sites, and the summation of $n$ is over $M$ cells.  For PBC case, the phase boundaries of the phase diagram is determined by the gap closing condition: $|t_{1R}/t_{2L}|=1$ or $|t_{1L}/t_{2R}|=1$, as shown in Fig.\ref{fig3}(a). For simplicity, all parameters  $t_{1L/1R}$ and $t_{2L/2R}$ are taken to be positive.

In the same framework we can obtain the analytical solution of model (\ref{GSSH}) \cite{supp}.
From the expression of $E$ in terms of $z_i$, it follows that $z_1$ and $z_2$ fulfill the constraint condition:
\begin{equation}\label{Sz1z2}
z_{1}z_{2}=\frac{t_{1R}t_{2R}}{t_{1L}t_{2L}} .
\end{equation}
Similarly, the boundary equation leads to
\begin{equation}
\begin{split}\label{Sqqz1}
&(z_{1}^{M+1}-z_{2}^{M+1})+\chi_1(z_{1}^{M}-z_{2}^{M})-\\
&\chi_2(z_{1}^{M-1}-z_{2}^{M-1})
- \chi_3(z_{1}-z_{2})=0
\end{split}
\end{equation}
with $\chi_1=\frac{t_{2R}t_{2L}-\delta _{R}\delta _{L}}{t_{1L}t_{2L}}$, $\chi_2=\frac{t_{1R}\delta _{R}\delta _{L}}{t_{1L}t_{2L}^{2}}$ and $\chi_3= \frac{\delta _{L}}{t_{2L}}+\frac{\delta _{R}%
}{t_{2R}}\left( \frac{t_{1R}t_{2R}}{t_{1L}t_{2L}}\right) ^{M} $.
Due to the constraint condition of Eq.(\ref{Sz1z2}), we can always set the solution as the form of Eq.(\ref{solution-set})
with $r=\sqrt{\frac{t_{1R}t_{2R}}{t_{1L}t_{2L}}}$. Then Eq.(\ref{Sqqz1}) becomes
\begin{equation}\label{Seq-theta}
\sin[(M+1)\theta]+\eta_1\sin[M\theta]-\eta_2\sin[(M-1)\theta]=\eta_3\sin[\theta],
\end{equation}
where $\eta_1=\frac{t_{2R}t_{2L}-\delta_R\delta_L}{\sqrt{t_{1R}t_{2R}t_{1L}t_{2L}}}$,   $\eta_2=\frac{\delta_R\delta_L}{t_{2R}t_{2L}}$ and $\eta_3=\frac{\delta _{L}}{t_{2L}} r ^{-M}+\frac{\delta _{R}}{t_{2R}} r ^{M}$.  The corresponding eigenvalue can be expressed as
$E=\pm\sqrt{2\sqrt{t_{1R}t_{2R}t_{1L}t_{2L}}\cos \theta+t_{1R}t_{1L}+t_{2R}t_{2L}}$ ,
where $\theta$ may take real or complex depending on the values of $\eta_1$,  $\eta_2$ and $\eta_3$.

When $\delta _{R}=\delta _{L}=0$, i.e., the OBC case,  we have $\eta_2=\eta_3=0$ and $\eta_1=\alpha$ with $\alpha=\sqrt{\frac{t_{2R}t_{2L}}{t_{1R}t_{1L}}}$.
While Eq.(\ref{Seq-theta}) has $M$ real solutions corresponding to bulk states when $\alpha<\alpha_c$, it has $M-1$ real solutions corresponding to bulk states and one complex solution ($\theta=\pi+i\varphi$) corresponding to edge states when $\alpha>\alpha_c$. In the thermodynamic limit,  $\alpha_c=1+\frac{1}{M } \rightarrow 1$, and thus the boundary of topological phase transition is given by $\alpha=1$, i.e., $t_{2R}t_{2L}=t_{1R}t_{1L}$ as shown in Fig.\ref{fig3}(b).

For the case with  $\delta _{L}=0$ and $\delta _{R} \neq 0$, we have $\eta_1=\alpha$, $\eta_2=0$ and $\eta_3=\frac{\delta _{R}}{t_{2R}} r ^{M}$.
In the thermodynamic limit we have $\eta_3\rightarrow 0$ as long as  $t_{1R}/t_{2L}<t_{1L}/t_{2R}$ ($r<1$), and the solutions of $\theta$ are identical to the OBC case. On the other hand, when $t_{1R}/t_{2L}>t_{1L}/t_{2R}$ ($r>1$), we have $\eta_3\rightarrow \infty$ for a finite $\delta _{R}/t_{2R}$ in the thermodynamic limit. It follows that  Eq.(\ref{Seq-theta}) has no real solutions but $M$ complex solutions, and we have
$ |z_1| \rightarrow 1$ and $|z_2|\rightarrow \frac{t_{1R}t_{2R}}{t_{1L}t_{2L}}$ for bulk states as $M \rightarrow \infty$. In this case, the spectrum in the thermodynamic limit approaches to the spectrum of system with PBC. Since the spectra in the regions of $r<1$ and $r>1$ take different forms, the phase diagram should display different behaviors in these regions as shown  in Fig.\ref{fig3}(c). While the phase boundary in the region of $r<1$ is similar to OBC case, it is similar to PBC case in the region of $r>1$.

For the other case with $\delta _{R}=0$  and $\delta _{L}\neq0$, we have $\eta_3=\frac{\delta _{L}}{t_{2L}} r ^{-M}$. In the large M limit, $\eta_3\rightarrow 0$ for $r>1$ and  $\eta_3\rightarrow \infty$ for $r<1$. Similarly, we can get the phase diagram as shown Fig.\ref{fig3}(d).  We note that the phase boundaries in Fig.\ref{fig3}(c) and Fig.\ref{fig3}(d) are determined by the gap closing conditions. In the shadow regions, there exist in-gap bound states. Also there are only left or right skin states in Fig.3(c) or Fig.3(d), in contrast to the OBC case.

In the presence of finite $\delta_L$ and $\delta_R$, 
Eq.(\ref{Seq-theta}) does not support real solutions in the large $M$ limit, and the spectrum shall approach the spectrum of the PBC case. Similar to the HN model, in the thermodynamic limit the NHSE is unstable to the perturbation with both  $\delta_L \neq 0$ and $\delta_R \neq 0$,  while it may exist in the finite size system.

{\it Conclusions and discussions.-} We exactly solved the non-Hermitian HN model and SSH model with GBCs and predicted the existence of NHSE beyond the OBC when one of the boundary hopping terms is absent.  Apart from this critical line on the boundary parameter space, the NHSE is unstable under tiny boundary perturbations and vanishes in the thermodynamic limit, whereas it may exist in a finite size system. We also applied our analytical results to explore the phase diagram of non-Hermitian SSH model under different boundary conditions and identified a novel phase diagram in the critical boundary line. Our exact solutions also provide an analytical method to determine finite-size generalized Brillouin zones.
We note that the fragility of NHSE under tiny boundary perturbations can be also found in other 1D nonreciprocal systems and even  higher-dimensional systems with NHSE \cite{GongJB,Nori}. Consider a n-dimensional nonreciprocal system which exhibits NHSE under OBC, if we add a boundary perturbation along one of directions and take PBC in the other $n-1$ directions, the higher-dimensional system can be mapped to a 1D nonreciprocal system with GBC by applying Fourier transformation. Then we can conclude the existence of fragility of NHSE by following similar calculations in 1D systems. We also give examples of two-dimensional (2D) models in the SM \cite{supp} and numerically confirm the fragility of NHSE in the large-size limit under tiny boundary perturbations along both x- and y-directions.

\begin{acknowledgments}
The work is supported by National Key
Research and Development Program of China (2016YFA0300600), NSFC under Grants No.11974413, and the Strategic Priority Research Program of Chinese Academy of Sciences under Grant No. XDB33000000.
\end{acknowledgments}

\onecolumngrid
\newpage
\renewcommand{\theequation}{S\arabic{equation}}
\renewcommand{\thefigure}{S\arabic{figure}}
\renewcommand{\thetable}{S\arabic{table}}
\setcounter{equation}{0}
\setcounter{figure}{0}
\setcounter{table}{0}

\begin{center}
    {\bf \large Supplemental Material for ``Exact solution of non-Hermitian systems with generalized boundary conditions: size-dependent boundary effect and fragility of skin effect" }
\end{center}

%
%
%
%
%
%
%

\section{Details for solutions of Hatano-Nelson model with generalized boundary conditions}

In this section of the supplementary material, we present some details for the analytical solutions of Hatano-Nelson model with generalized boundary conditions. To keep consistence with the main text, here we take parameters as $\mathrm{sgn}[t_{R}t_{L}]=1$, $\mathrm{sgn}[\delta_{R}t_{R}]\neq-1$ and $\mathrm{sgn}[\delta_{L}t_{L}]\neq-1$.

As shown in the main text, the expression of eigenvalue in terms of $z_i$ can be written as:
\begin{eqnarray}\label{XHEz1}
E &=&\frac{t_{R}}{z_{i}}+t_{L}z_{i},
\end{eqnarray}
and the general ansatz of wave function can be described as
\begin{equation}
\Psi=c_{1}\Psi _{1}+c_{2}\Psi _{2} = (\psi _{1},\psi _{2},\cdots ,\psi _{N})^{T},~~~ \label{XHWave}
\end{equation}
where $\psi _{n}=\sum_{i=1}^{2}(c_{i}z_{i}^{n})=c_{1}z_{1}^{n}+c_{2}z_{2}^{n}$ with $n=1,2,\cdots ,N$.

For nontrivial solutions for $(c_1, c_2)$, i.e.,
$c_1 = 0$ and $c_2 = 0$ cannot be satisfied simultaneously, $z_1, z_2$ should satisfy the following two conditions
\begin{equation}
z_{1}z_{2}=\frac{t_R}{t_L},
\end{equation}
\begin{equation}\label{XHqqz1}
(z_{1}^{N+1}-z_{2}^{N+1})-\frac{\delta _{R}\delta _{L}}{t_{L}^{2}}%
(z_{1}^{N-1}-z_{2}^{N-1})-\left[ \frac{\delta _{L}}{t_{L}}+\frac{\delta _{R}}{t_{R}}\left( \frac{t_{R}}{t_{L}}\right) ^{N}\right] (z_{1}-z_{2})=0 .
\end{equation}
We can always set the solution as
\begin{equation}
z_{1}=re^{i\theta}, ~~~ z_{2}=re^{-i\theta}
\end{equation}
with $r=\sqrt{\frac{t_{R}}{t_{L}}}$, then Eq.(\ref{XHqqz1})  becomes
\begin{equation}\label{XHeq-theta}
\sin[(N+1)\theta]-\eta_1\sin[(N-1)\theta]-\eta_2\sin[\theta]=0,
\end{equation}
where $\eta_1=\frac{\delta_R\delta_L}{t_Rt_L}$ and $\eta_2=\frac{\delta _{L}}{t_{L}} r ^{-N}+\frac{\delta _{R}}{t_{R}} r ^{N}$
.  The corresponding eigenvalue can be expressed as
\begin{equation}
E=2\sqrt{t_{R}t_{L}}\cos \theta . \label{XHspectrum1}
\end{equation}
The solutions $\theta$ of Eq.(\ref{XHeq-theta}) may take real or complex depending on the values of $\eta_1$ and $\eta_2$.

Under the open boundary condition (OBC), since $\delta_R=0$ and $\delta_L=0$, we have $\eta_1=\eta_2=0$, and thus Eq.(\ref{XHeq-theta}) reduces to
\begin{equation}\label{XHeq-theta-OBC}
\sin[(N+1)\theta]=0.
\end{equation}
The solution of above equation takes $N$ real roots: $\theta=\frac{m\pi}{N+1}$ ($m=1,\cdots, N$).

When either $\delta_R=0$ or $\delta_L=0$, we have $\eta_1=0$, and Eq.(\ref{XHeq-theta}) becomes
\begin{equation}\label{XHeq-noeta1}
\sin[(N+1)\theta]-\eta_2 \sin[\theta]=0
\end{equation}
with $\eta_2=\frac{\delta _{L}}{t_{L}} r ^{-N}$  or $\eta_2 = \frac{\delta _{R}}{t_{R}} r ^{N}$. If $\eta_2<\eta_c$, Eq.(\ref{XHeq-noeta1}) has $N$ real roots.
When $\eta_2>\eta_c$, $\theta$ begins to take complex root,
where $\eta_c=1$ for $N=4j$, $\eta_c=1/\cos [\frac{\pi}{N+1}]$ for $N=4j+2$, and $\eta_c=1/\cos [\frac{\pi}{2(N+1)}]$ for $N=odd$, here $j=1,2,\cdots$. It is clear $\eta_c$ always approaches $1$ as $N \rightarrow \infty$. When $\eta_2> N+1$,  $\theta$ has no real roots.

When $\eta_2=\eta_c$, the roots of Eq.(\ref{XHeq-noeta1}) are all real, and some roots are degenerate. For example, when $\eta_2=\eta_c=1$ for $N=4j$, there are $N-1$ real roots and no complex roots. In this case, Eq.(\ref{XHeq-noeta1}) becomes
\begin{equation}
\cos[\frac{(N+2)\theta}{2}]  \sin[\frac{N\theta}{2}]=0.
\end{equation}
The solution of above equation is $\theta=\frac{2m+1}{N+2}\pi$ with $m=0, \cdots, N/2$ and  $\theta=\frac{2 n}{N}\pi$ with $n=1, \cdots, N/2-1$.  The number of real solution is
$N-1$, because the degeneracy of solution of $\theta=\frac{\pi}{2}$ is 2.

Consider the case of $r>1$ ($|t_R|>|t_L|$), when $\delta_R=0$ we have $\eta_2=\frac{\delta _{L}}{t_{L}} r ^{-N}$. We have always $\eta_2<\eta_c$ unless $\frac{\delta _{L}}{t_{L}}>\eta_cr^{N}$. For a fixed $\delta _{L}$, it is clear  $\eta_2 \rightarrow 0$ when $N \rightarrow \infty$, i.e., in the large N limit, the system is identical to the OBC. On the other hand,  when $\delta_L=0$, we have $\eta_2 = \frac{\delta _{R}}{t_{R}} r ^{N}$. Therefore, if $\frac{\delta _{R}}{t_{R}} < \eta_c r ^{-N}$,  Eq.(\ref{XHeq-noeta1}) has $N$ real roots, and the system exhibits non-Hermitian skin effect. With the increase of $N$, the region with non-Hermitian skin effect becomes narrower with the boundary $\eta_c r ^{-N}$ decreasing exponentially to zero. It is clear that non-Hermitian skin effect only occurs in the axis of $\frac{\delta_{R}}{t_{R}}=0$ in the thermodynamic limit.
For the case of $r<1$ ($|t_R|<|t_L|$), we can make similar analysis. It follows that  non-Hermitian skin effect only occurs in the axis of $\frac{\delta _{L}}{t_{L}}=0$ in the thermodynamic limit.

Next we consider the general case with arbitrary $\delta_L$ and $\delta_R$.
In the region of $0\leq\frac{\delta_R}{t_R}, \frac{\delta_L}{t_L}\leq1$, we have $0\leq\eta_1\leq 1$ and $\eta_2\geq0$. Eq.(\ref{XHeq-theta}) can be rewritten as
\begin{equation}
f_1=f_2
\end{equation}
with
\begin{equation}
f_1=(1+\eta_1)\cos[N\theta]\sin[\theta]+(1-\eta_1)\sin[N\theta]\cos[\theta], ~~~~~~f_2=\eta_2\sin[\theta].
\end{equation}
The outer contour of $(1+\eta_1)\cos[N\theta]\sin[\theta]$ obeys $\pm(1+\eta_1)\sin[\theta]$, and the outer contour of $(1-\eta_1)\sin[N\theta]\cos[\theta]$ obeys $\pm(1-\eta_1)\cos[\theta]$. Thus, the outer contour of $f_1$ is larger than $\max\{ \pm(1+\eta_1)\sin[\theta], \pm(1-\eta_1)\cos[\theta]\}$, and  the outer contour of $f_1$ is approximately equal to  $\pm(1+\eta_1)\sin[\theta]$ when $\theta$  is near $\pi/2$, but $\pm(1-\eta_1)\cos[\theta]$ when $\theta$ is near $0,\pi$. The intersections of $f_1$ and $f_2$ determine the real solutions of $\theta$, there are $N$ real $\theta$ when the outer contour of $f_1$ is larger than that of $f_2$. Because the first intersection to disappear as $\eta_2$ increases is near $\theta=\pi/2$, the condition of $N$ real solutions $\theta$ is approximately given by $\eta_2<1+\eta_1$.

When $\eta_2>1+\eta_1$, with the increase in $\eta_2$, the intersections of $f_1$ and $f_2$ will gradually disappear from the points near $\theta=\pi/2$ to near $\theta=0,\pi$, and the number of real solutions gradually decreases. Because the last point to disappear is near $\theta=0$, the condition of no real solution is determined by $f_2'(\theta=0)>f_1'(\theta=0)$, where $f_i' = \frac{\partial f_i}{\partial \theta}$. It gives rise to $\eta_2>N+1-\eta_1(N-1)$.

In particularly, if $\frac{\delta _{R}}{t_{R}}=\frac{\delta _{L}}{t_{L}}=\Gamma$, we have $\eta_1=\Gamma^2$ and $\eta_2=\Gamma(r ^{N}+r ^{-N})$. When $\Gamma=r^{\pm N}$, it is easy to check $\eta_2=1+\eta_1$. In the region of $0\leq\Gamma\leq1$, we have $\Gamma_c=r ^{N}$ for $r<1$ and $\Gamma_c=r ^{-N}$ for $r>1$. When $0<\Gamma<\Gamma_c$, all solutions take real roots and the finite-size skin effect exists. And the condition of $\eta_2=N+1-\eta_1(N-1)$ is fulfilled when $\Gamma=\frac{-(r ^{N}+r ^{-N})+\sqrt{(r ^{N}+r ^{-N})^2+4(N^2-1)}}{2(N-1)}$.

When $\eta_2<1+\eta_1$, there are $N$ real solutions for Eq.(\ref{XHeq-theta}), thus $|z_{1/2}|=r$ and the corresponding eigenvalues are real.
When $1+\eta_1<\eta_2<N+1-\eta_1(N-1)$, there are some real solutions and some complex solutions for Eq.(\ref{XHeq-theta}), thus the eigenvalues are not all real and  some complex eigenvalues appear. When $\eta_2>N+1-\eta_1(N-1)$, there is no real solution and $N$ complex solutions occur for Eq.(\ref{XHeq-theta}). Consequently, almost all energies are not real but complex except for the ones corresponding to the angle of $z_i$ being $0,\pi$. In the thermodynamic limit, $\eta_2>N+1-\eta_1(N-1)$ always holds true for cases of $\delta_R=0$ (a fixed $\delta_L$) with $r<1$, $\delta_L=0$ (a fixed $\delta_R$) with $r>1$, and fixed $\delta_R\neq0$ and $\delta_L\neq0$, because $\eta_2$ increases exponentially with $N$.

For the case of the existence of $N$ complex solutions for Eq.(\ref{XHeq-theta}), we have $|z_{1}|\neq|z_{2}|$. If we set $\theta=\theta_R+i\theta_I$, Eq.(\ref{XHeq-theta}) can be rewritten as
\begin{equation}
e^{i(N+1)(\theta_R+i\theta_I)}-e^{-i(N+1)(\theta_R+i\theta_I)}-\eta_1\left(e^{i(N-1)(\theta_R+i\theta_I)}-e^{-i(N-1)(\theta_R+i\theta_I)}\right)=\eta_2\left(e^{i(\theta_R+i\theta_I)}-e^{-i(\theta_R+i\theta_I)}\right).
\end{equation}
In the thermodynamic limit, after we ignore the very small terms which exponentially approaching zero, the equation above becomes
\begin{equation}
e^{(N+1)\theta_I}e^{-i(N+1)\theta_R}-\eta_1e^{(N-1)\theta_I}e^{-i(N-1)\theta_R}=\eta_2e^{\theta_I}e^{-i\theta_R} \left(1-e^{-2\theta_I}e^{i2\theta_R}\right),
\end{equation}
where we have assumed $\theta_I>0$. Therefore, we have
\begin{equation}\label{XHeq-theta2}
e^{N\theta_I}=\eta_2a_1e^{iN\theta_R}
\end{equation}
with
\begin{equation}
a_1=\frac{1-e^{-2\theta_I}e^{i2\theta_R}}{1-\eta_1e^{-2\theta_I}e^{i2\theta_R}}.
\end{equation}
When we focus on the real part and imaginary part of Eq.(\ref{XHeq-theta2}), we can obtain
\begin{equation}
\begin{split}
e^{\theta_I}&=\sqrt[N]{\eta_2}\sqrt[N]{|a_1|},\\
\theta_R&=\frac{2m\pi-\theta_{a_1}}{N}~~~(m=1,2,\cdots,N),
\end{split}
\end{equation}
where $\theta_{a_1}$ is the angle of $a_1$, i.e., $a_1 =|a_1| e^{i\theta_{a_1}}$. 

\begin{figure*}[tbp]
\includegraphics[width=1.0\textwidth]{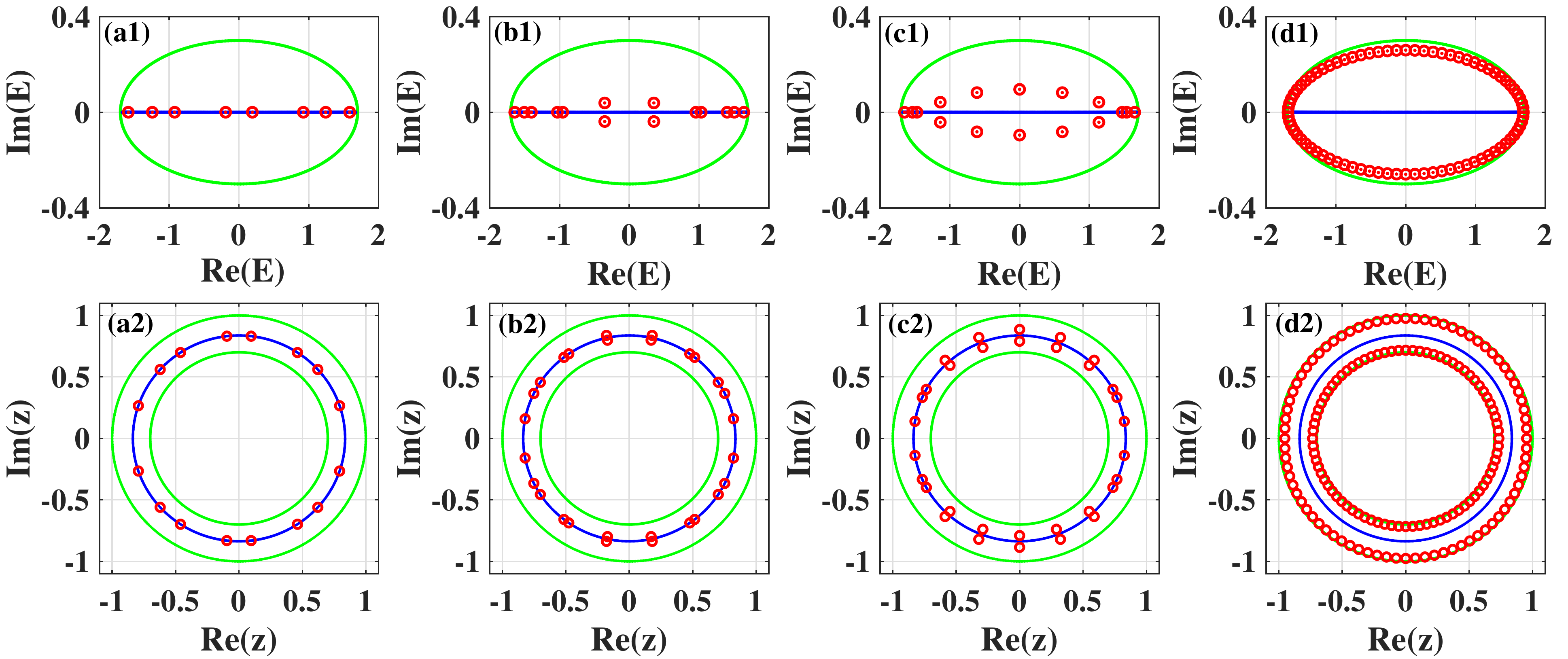}
\caption{(a1,b1,c1,d1) Eigenvalues of Hatano-Nelson model for $N=8,14,16$ and $80$, respectively. The analytical results (red circles) are in exact agreement with the numerical results (red dots). The green and blue curves represent spectra corresponding  to the PBC case and OBC case in the thermodynamic limit, respectively. (a2,b2,c2,d2) The finite-size generalized Brillouin zones $z$ described by red circles for $N=8,14,16,80$, respectively. The green and blue curves represent the BZs for PBC case and the GBZs for OBC case in the thermodynamic limit, respectively. Common parameters: $t_{L}=1, t_{R}=0.7, \frac{\delta_{L}}{t_{L}}=0.09, \frac{\delta_{R}}{t_{R}}=0.8$.}%
\label{Xfig1}
\end{figure*}
\begin{figure*}[tbp]
\includegraphics[width=1.0\textwidth]{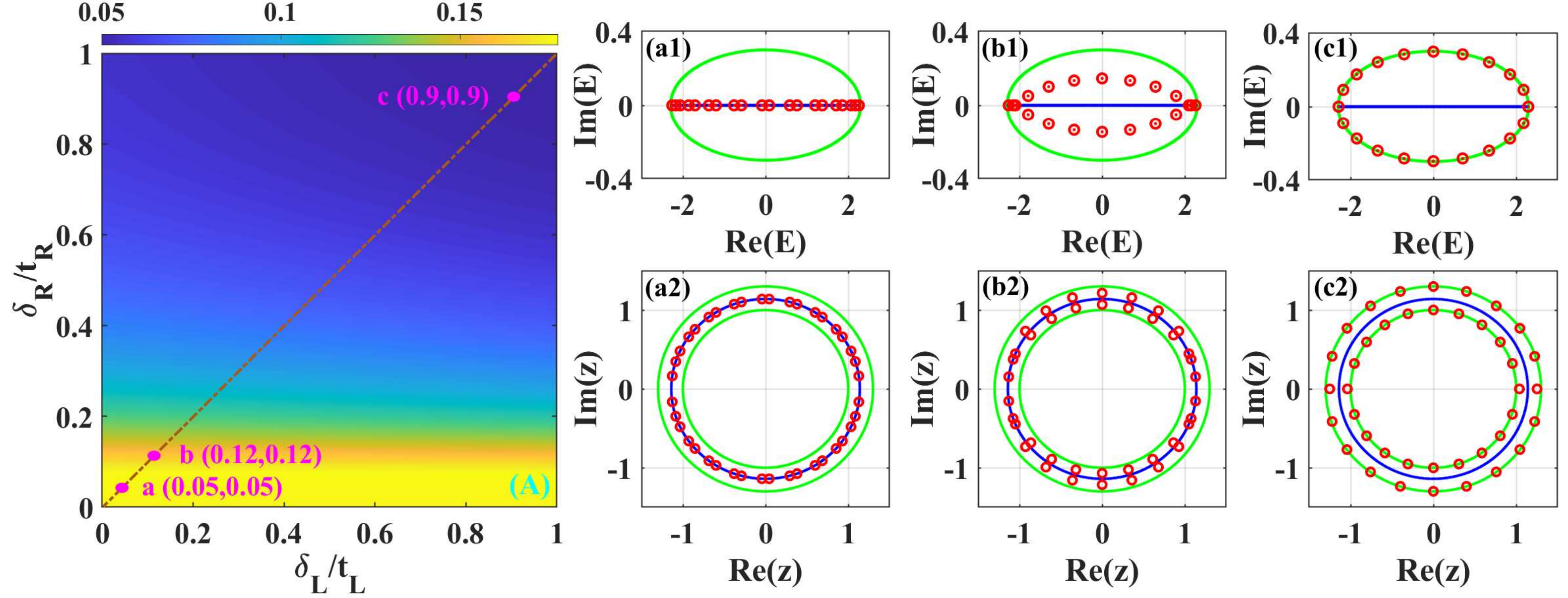}
\caption{(A) $\overline{\mathrm{IPR}}$ in the parameter space of $\delta_L/t_L$ and  $\delta_R/t_R$ for Hatano-Nelson model with $N=20, t_{L}=1, t_{R}=1.3$. (a1,b1,c1) Eigenvalues corresponding to dots 'a,b,c' in (A), respectively. The analytical results (red circles) are in exact agreement with the numerical results (red dots). The green and blue curves represent energy spectra corresponding PBC case and OBC case in the thermodynamic limit, respectively. (a2,b2,c2) The finite-size generalized Brillouin zones $z$ described by red circles corresponding dots 'a,b,c' in (A), respectively. The green and blue curves represent the BZs for PBC case and the GBZs for OBC case in the thermodynamic limit, respectively.}%
\label{Xfig2}
\end{figure*}

If $r>1$, we have $\eta_2\approx \frac{\delta _{R}}{t_{R}} r ^{N}$, and $\sqrt[N]{|a_1|}\rightarrow 1$ due to $|a_1|$ is a finite number. So we have $e^{\theta_I}=r\sqrt[N]{\frac{\delta_R}{t_R}}$. Since
\begin{equation}
z_1=re^{-\theta_I}e^{i\theta_R},~~~  z_2=re^{\theta_I}e^{-i\theta_R},
\end{equation}
then we have
\begin{equation}
z_1\approx\sqrt[N]{\frac{t_R}{\delta_R}}e^{i\frac{2m\pi-\theta_{a_1}}{N}},~~~ z_2\approx\frac{t_R}{t_L}\sqrt[N]{\frac{\delta_R}{t_R}}e^{-i\frac{2m\pi-\theta_{a_1}}{N}}.
\end{equation}
When $N\rightarrow \infty$, $|z_1|\approx  \sqrt[N]{\frac{t_R}{\delta_R}} \rightarrow 1$ and $|z_2|\approx \frac{t_R}{t_L}\sqrt[N]{\frac{\delta_R}{t_R}} \rightarrow \frac{t_R}{t_L}$, therefore the spectrum  approaches to the periodic spectrum in the thermodynamic limit.
If $r<1$, we have $\eta_2\approx \frac{\delta _{L}}{t_{L}} r ^{-N}$, $\sqrt[N]{|a_1|}\rightarrow 1$ and $e^{\theta_I}=r^{-1}\sqrt[N]{\frac{\delta_L}{t_L}}$, giving rise to
\begin{equation}
z_1\approx\frac{t_R}{t_L}\sqrt[N]{\frac{t_L}{\delta_L}}e^{i\frac{2m\pi-\theta_{a_1}}{N}},~~~ z_2\approx\sqrt[N]{\frac{\delta_L}{t_L}}e^{-i\frac{2m\pi-\theta_{a_1}}{N}}.
\end{equation}
Due to $|z_2|\approx \sqrt[N]{\frac{\delta_L}{t_L}} \rightarrow 1$ and $|z_1|\approx  \frac{t_R}{t_L} \sqrt[N]{\frac{t_L}{\delta_L}}\rightarrow \frac{t_R}{t_L}$ when $N \rightarrow \infty$, the spectrum approaches to the periodic spectrum in the thermodynamic limit.
Therefore, the case of $N$ complex solutions in the thermodynamic limit is similar to the PBC case.

In Fig.\ref{Xfig1}, we display eigenvalues and the finite-size generalized Brillouin zones $z$ of the Hatano-Nelson model with the same parameters for different lattice sizes. For $N=8$ in (a1,a2), all $|z_{1/2}|=r$ located at the generalized Brillouin zone (GBZ) for the OBC case, and eigenvalues are all real. For $N=14$ and $16$ in (b1,b2,c1,c2), some $|z_{1/2}|=r$ located at the GBZ for the OBC case, while some $|z_{1/2}|\neq r$. Eigenvalues corresponding to $|z_{1/2}|=r$ are real and the others are complex. For $N=80$ in (d1,d2), all $|z_{1/2}|\rightarrow 1$ approximately located at Brillouin zones (BZs) for the PBC case, and eigenvalues are in close proximity to the periodic spectrum.

In Fig.\ref{Xfig2}(A), we display $\overline{\mathrm{IPR}}$ in the parameter space of $\delta_L/t_L$ and  $\delta_R/t_R$  for Hatano-Nelson model with $N=20, t_{L}=1$ and $t_{R}=1.3$. The eigenstates in the yellow region are similar to the OBC case, and the corresponding eigenvalues are real as displayed in Fig.\ref{Xfig2}(a1) and the corresponding $|z_{1/2}|=r$ located at the GBZ for the OBC case as displayed in Fig.\ref{Xfig2}(a2). In the transition zone, some eigenvalues corresponding $|z_{1/2}|=r$ are real and others are complex as displayed in Figs.\ref{Xfig2}(b1,b2). The eigenvalues in the blue region are similar to the PBC case as displayed in Fig.\ref{Xfig2}(c1), and all $|z_{1/2}|\rightarrow 1, |z_{2/1}|\rightarrow \frac{t_R}{t_L}$ approximately located at BZs for the PBC case as displayed in Fig.\ref{Xfig2}(c2). According to our analytical result, the finite-size skin effect exists when $\Gamma=\delta_L/t_L=\delta_R/t_R$ is smaller than $\Gamma_c = r^{-20}\approx 0.07$ along the diagonal line of the parameter space.

\section{Spectral flow of Hatano-Nelson model with generalized boundary conditions}
\begin{figure*}[tbp]
\includegraphics[width=1.0\textwidth]{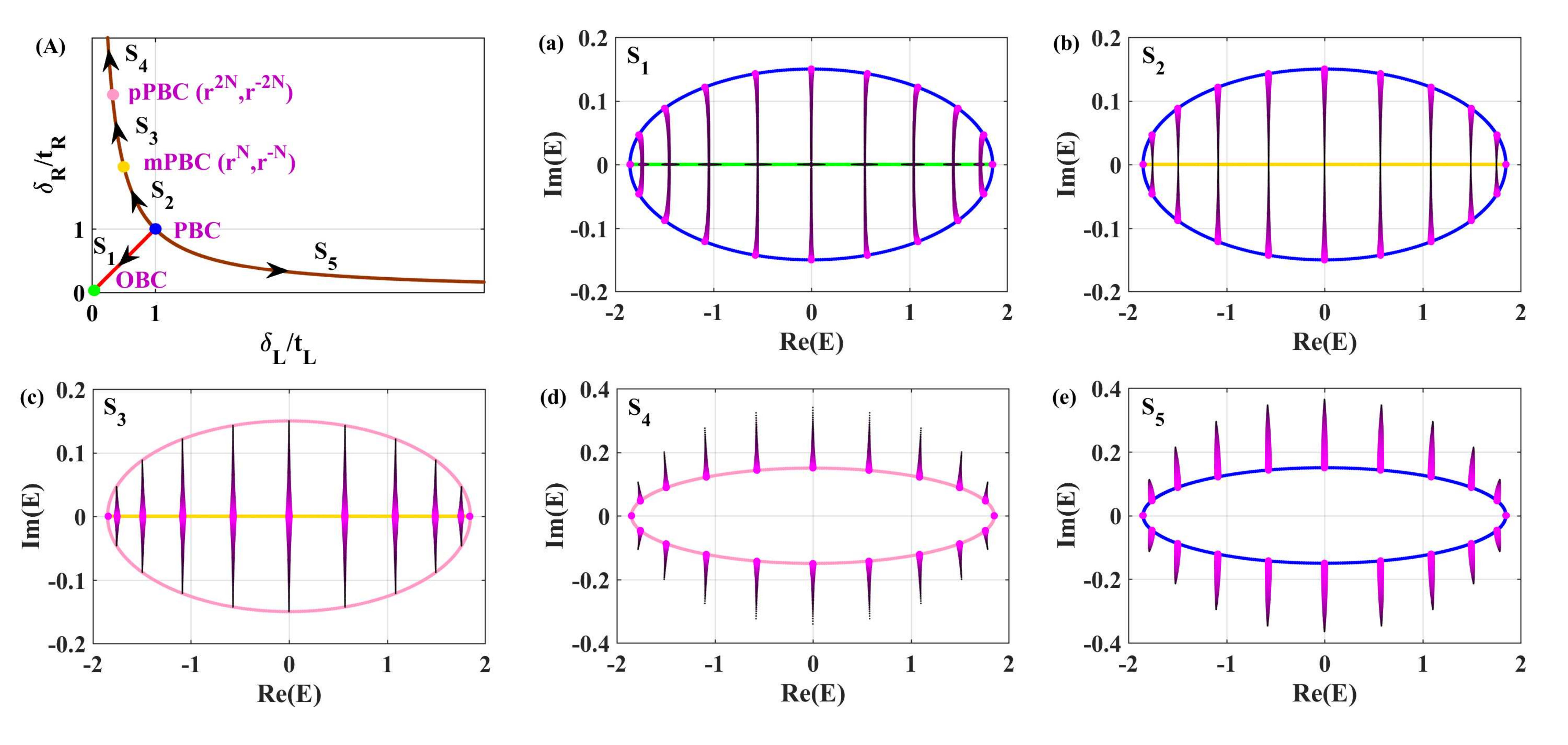}
\caption{(A) The schematic diagram of some routes for spectral flow on the parameter space of $(\frac{\delta_{L}}{t_{L}}, \frac{\delta_{R}}{t_{R}})$ for the HN model with $\frac{t_{R}}{t_{L}}<1$. (a) The spectral flow (magenta-black dots) along the route $\mathrm{S_1}$ from $\mathrm{PBC}$ to $\mathrm{OBC}$. (b) The spectral flow (magenta-black dots) along the route $\mathrm{S_2}$ from $\mathrm{PBC}$ to $\mathrm{mPBC}$. (c) The spectral flow (magenta-black dots) along the route $\mathrm{S_3}$ from $\mathrm{mPBC}$ to $\mathrm{pPBC}$. (d) The spectral flow (magenta-black dots) along the route $\mathrm{S_4}$ from $\mathrm{pPBC}$ to $(\frac{\delta_{L}}{t_{L}}, \frac{\delta_{R}}{t_{R}})=(\mu,\frac{1}{\mu})$ with $\mu=0.005$. (e) The spectral flow (magenta-black dots) along the route $\mathrm{S_4}$ from $\mathrm{PBC}$ to $(\frac{\delta_{L}}{t_{L}}, \frac{\delta_{R}}{t_{R}})=(\mu,\frac{1}{\mu})$ with $\mu=10$. In (a-e), The blue, green, orange and pink curves represent energy spectra under PBC, OBC, mPBC and pPBC in the thermodynamic limit, respectively. Common parameters: $t_L = 1, t_R = 0.85, N=20$.}%
\label{Xfig21}
\end{figure*}
\begin{figure*}[tbp]
\includegraphics[width=1.0\textwidth]{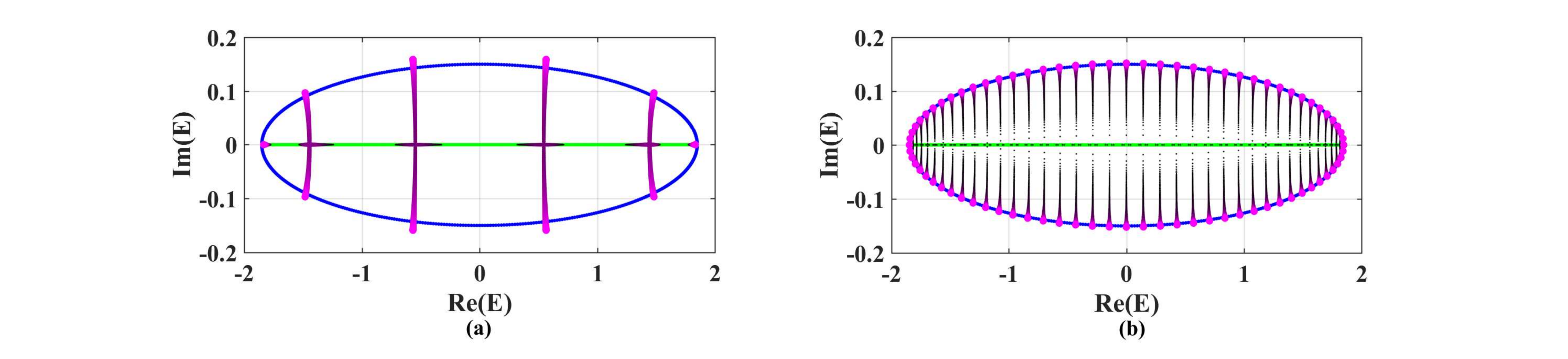}
\caption{The spectral flow (black-magenta dots) along the line $(\frac{\delta_{L}}{t_{L}}, \frac{\delta_{R}}{t_{R}})=(x,0.8)$ with changing $x$ from 0 to 1 for Hatano-Nelson model with generalized boundary condition. (a) $N=10$. (b) $N=80$. The blue and green curves represent energy spectra under PBC and OBC in the thermodynamic limit, respectively. Common parameters: $t_L = 1, t_R = 0.85$.
}%
\label{Xfig22}
\end{figure*}

In order to see clearly how the energy spectrum varies continuously with the change of boundary parameters, we show the spectral flow along some routes on the parameter space of $(\frac{\delta_{L}}{t_{L}}, \frac{\delta_{R}}{t_{R}})$ for the Hatano-Nelson model with generalized boundary conditions in this section.
We first discuss two cases as schematically displayed in Fig.\ref{Xfig21}(A). The first case is along the route $S_1$ which interpolates between PBC and OBC continuously, and the second case is along the route described by  $\frac{t_{R}}{\delta _{R}}=\frac{\delta _{L}}{t_{L}}$ as highlighted by the brown line in Fig.\ref{Xfig21}(A), which corresponds to the case of $(c_i\neq0, c_j=0) (i,j=1,2)$. Particularly, the points marked by PBC, mPBC and pPBC are all located on this line, corresponding to the boundary conditions of PBC, mPBC (modified PBC) \cite{ImuraSM} and pPBC (pseudo-PBC), respectively.  Before showing the spectral flow, we first give analytical expressions of energy spectra for these four particular boundary conditions, i.e., PBC $(\frac{\delta _{L}}{t_{L}}=\frac{\delta _{R}}{t_{R}}=1)$, OBC $(\frac{\delta _{L}}{t_{L}}=\frac{\delta _{R}}{t_{R}}=0)$, mPBC $(\frac{\delta _{L}}{t_{L}}=r^N,\frac{\delta _{R}}{t_{R}}=r^{-N})$, pPBC $ (\frac{\delta _{L}}{t_{L}}=r^{2N},\frac{\delta _{R}}{t_{R}}=r^{-2N})$ as marked in Fig.\ref{Xfig21}(A). While the spectrum under PBC is given by
\begin{equation}
E=(t_{L}+t_{R})\cos (\theta)+i(t_{L}-t_{R})\sin (\theta ), ~~~~ \theta=\frac{2m\pi}{N}
\end{equation}
 with $m=1,2,\cdots ,N$, the spectrum under OBC reads as
\begin{equation}
 E=2\sqrt{t_{R}t_{L}}\cos \theta, ~~~~\theta=\frac{m\pi }{N+1}
\end{equation}
 with $m=1,2,\cdots ,N$. The energy spectrum under mPBC is given by
\begin{equation}
 E=2 \sqrt{t_L t_R} \cos \theta, ~~~~\theta=\frac{2m\pi}{N}
\end{equation}
 with $m=1,2,\cdots ,N$. Here, we notice that the spectrum under mPBC are also real and similar to the spectrum under OBC. From the main text, it is known that the eigenstates corresponding to a given $\mu$ is given by
\begin{equation}
\Psi= \left(\sqrt[N]{\mu }e^{i\theta},\left( \sqrt[N]{\mu }e^{i\theta} \right)^{2},\cdots,\left( \sqrt[N]{\mu }e^{i\theta} \right)^{N} \right)^{T},
\end{equation}
and therefore the wave function under mPBC takes the following form
\begin{equation}
\Psi= \left(r e^{i\theta}, r^2 e^{2i\theta},\cdots, r^{N} e^{iN \theta}  \right)^{T},
\end{equation}
which is also similar to the wave function under OBC and exhibits non-Hermitian skin effect.
Therefore, the energy spectrum under mPBC can reconstruct energy spectrum under OBC in the thermodynamic limit\cite{ImuraSM}. The energy spectrum under pPBC can be expressed as
\begin{equation}
E=(t_{L}+t_{R})\cos (\theta)-i(t_{L}-t_{R})\sin (\theta ), ~~~~ \theta=\frac{2m\pi}{N}
\end{equation}
with $m=1,2,\cdots ,N$. By noticing that the values $\theta$ appear always in pairs of $(\theta, -\theta)$ except the case of $\theta=0,\pi (\sin[0]=\sin[\pi]=0)$, we can see that the spectrum under pPBC are the same as the spectrum under PBC. However, the corresponding wave function given by
\begin{equation}
\Psi= \left(r^2 e^{i\theta}, r^4 e^{2i\theta},\cdots, r^{2N} e^{iN \theta}  \right)^{T}
\end{equation}
exhibits non-Hermitian skin effect. The system under pPBC has obviously different behavior from the PBC case, even they share the same spectrum structures. This is why we call such a boundary condition as pPBC.

In Fig.\ref{Xfig21}(A), we schematically mark points corresponding to mPBC  and pPBC for the case of $r<1$ and label routes $S_1, S_2, S_3, S_4, S_5$. The route $S_1$ interpolates between PBC and OBC when the parameter $\frac{\delta _{L}}{t_{L}}=\frac{\delta _{R}}{t_{R}}$ changes from $1$ to $0$. This route has been used to study the spectral flow from PBC to OBC numerically \cite{LeeCHSM}. The routes $S_2, S_3, S_4, S_5$ all belong to the case of $\frac{t_{R}}{\delta _{R}}=\frac{\delta _{L}}{t_{L}}=\mu$.
The route $S_2$ interpolates between PBC and mPBC and is characterized by parameters $(\mu,\frac{1}{\mu})$ with changing $\mu$ from $1$ to $r^{N}$. It is worth pointing out that the spectral flow along route $S_2$ for our system is just the same as spectral flow for the system $H_{\kappa}=H(k+i\kappa)$ with an imaginary flux $\kappa$  changing from $0$ to a critical value $\ln(r)$  studied in Ref. \cite{LeeCHSM}. By performing a gauge transform $H_{\kappa}\rightarrow V^{-1}H_{\kappa}V$ $(V=\mathrm{diag}\{e^{-\kappa},e^{-2\kappa},\cdots,e^{-N\kappa}\})$ without changing energy spectra, the Hamiltonian $H_{\kappa}=H(k+i\kappa)$ can be transformed into our system with boundary conditions $(e^{N\kappa},e^{-N\kappa})$, which is corresponding to our special case of $\frac{t_{R}}{\delta _{R}}=\frac{\delta _{L}}{t_{L}}=\mu$. When $\kappa=0$, the Hamiltonian after transformation is just our system with the boundary condition of PBC $(1,1)$. When $\kappa=\ln(r)$, the Hamiltonian after transformation is just our system with the boundary condition of mPBC $(r^N,r^{-N})$, whose energy spectra can reconstruct the spectra with OBC in the thermodynamic limit.
The route $S_3$ interpolates between mPBC and pPBC  and is characterized by parameters $(\mu,\frac{1}{\mu})$ with changing $\mu$ from $r^{N}$ to $r^{2N}$. The route $S_4$ is characterized by parameters $(\mu,\frac{1}{\mu})$ with changing $\mu$ from $r^{2N}$ to a number heading towards $0$. The route $S_5$ is characterized by parameters $(\mu,\frac{1}{\mu})$ with changing $\mu$ from $1$ to a number heading towards $\infty$.
In Figs.\ref{Xfig21}(a-e), we present the spectral flow (magenta-black dots) along the routs $\mathrm{S_1}, \mathrm{S_2}, \mathrm{S_3}, \mathrm{S_4}, \mathrm{S_5}$, and the spectral flow are consistent with our above analysis.

As a supplement, we also show spectral flow along another route on the parameter space of $(\frac{\delta_{L}}{t_{L}}, \frac{\delta_{R}}{t_{R}})=(x,0.8)$ with changing $x$ from $0$ to $1$ for HN model in Fig.\ref{Xfig22}. In Fig.\ref{Xfig22}(a), we plot the spectral flow for $N=10$, the parameters are the same as those of Fig.1(A) in the main text. When $x=1$, the spectrum for $N=10$ is not exactly equal to the spectrum under PBC, while for the system with $N=80$ and $x=1$, the spectrum  is very close to that under PBC, as plotted in Fig.\ref{Xfig22}(b). This is consistent with the analytical analysis in the main text that the spectrum approaches to the periodic spectrum in the thermodynamic limit.

\section{Analytical solutions of the non-Hermitian Su-Schrieffer-Heeger model with generalized boundary conditions}
Here we give the details for the derivation of the exact solution of the 1D two-band non-Hermitian Su-Schrieffer-Heeger model, with its Hamiltonian given by
\begin{equation}
\hat{H}=\sum\limits_{n=1}^{M}[ t_{1L}\hat{c}_{nA}^{\dag
}\hat{c}_{nB}+t_{1R}\hat{c}_{nB}^{\dag }\hat{c}_{nA}]+\sum\limits_{n=1}^{M-1}[t_{2R}\hat{c}_{(n+1)A}^{\dag
}\hat{c}_{nB}+t_{2L}\hat{c}_{nB}^{\dag }\hat{c}_{(n+1)A}]
+\delta _{R}\hat{c}_{1A}^{\dag
}\hat{c}_{MB}+\delta _{L}\hat{c}_{MB}^{\dag }\hat{c}_{1A},
\end{equation}
where $t_{1L/1R}$ and $t_{2L/2R}$ are imbalanced hopping term between intracell sites and intercell sites, and $M$ is the number of cells. Here, we also focus on the situation with all parameters  $t_{1L/1R}$ and $t_{2L/2R}$ taking positive.

The corresponding eigenvalue equation can be written as $\hat{H}|\Psi\rangle=E|\Psi\rangle$, where $|\Psi\rangle=\sum_n (\psi_{i,A} \hat{c}_{nA}^{\dag } + \psi_{i,B} \hat{c}_{nB}^{\dag } ) |0\rangle$. For convenience, we also
denote $\Psi = (\psi_{1A},\psi_{1B},\psi_{2A},\cdots,\psi_{MB})^T$. The above eigenvalue equations consist of a series of equations, including bulk equations as follows
\begin{eqnarray}
t_{1R}\psi _{sA}-E\psi _{sB}+t_{2L}\psi _{(s+1)A} &=&0,\label{XSbk1}\\
t_{2R}\psi _{sB}-E\psi _{(s+1)A}+t_{1L}\psi _{(s+1)B} &=&0\label{XSbk2}
\end{eqnarray}
with $s=1,\cdots ,M-1$, and the boundary equations given by $-E\psi _{1A}+t_{1L}\psi _{1B}+\delta _{R}\psi _{MB}=0$ and $\delta _{L}\psi _{1A}+t_{1R}\psi _{MA}-E\psi _{MB}=0$. By comparing the above two equations with Eq.(\ref{XSbk1},\ref{XSbk2}), they are equivalent to the following boundary conditions
\begin{eqnarray}
t_{2R}\psi _{0B} &=&\delta _{R}\psi _{MB}, \label{XSbd3}\\
\delta _{L}\psi _{1A}&=&t_{2L}\psi _{(M+1)A}. \label{XSbd4}
\end{eqnarray}

Due to spatial translational property from bulk equations, we set the ansatz of wave function $\Psi _{i}$ which satisfies the bulk equations Eq.(\ref{XSbk1},\ref{XSbk2}) as follows
\begin{equation}\label{XSFii}
\Psi _{i}=(z_{i}\phi _{A}^{(i)},z_{i}\phi _{B}^{(i)},z_{i}^{2}\phi
_{A}^{(i)},z_{i}^{2}\phi _{B}^{(i)},\cdots ,z_{i}^{M}\phi
_{A}^{(i)},z_{i}^{M}\phi _{B}^{(i)})^{T}.
\end{equation}
By inserting  Eq.(\ref{XSFii}) into the bulk equation Eq.(\ref{XSbk1},\ref{XSbk2}), we obtain the expression of eigenvalue in terms of $z_i$:
\begin{eqnarray}\label{XSEz1}
E &=& \pm \sqrt{\frac{t_{1R}t_{2R}}{z_{i}}%
+t_{1L}t_{2L}z_{i}+t_{1L}t_{1R}+t_{2L}t_{2R}},
\end{eqnarray}
and the relation between $\phi _{A}^{(i)}$ and $\phi _{B}^{(i)}$ as follows
\begin{eqnarray}\label{XSFAB}
\phi _{A}^{(i)}=\frac{E}{(t_{1R}+t_{2L}z_{i})}\phi _{B}^{(i)}=\frac{(t_{2R}+t_{1L}z_{i})}{Ez_{i}}\phi _{B}^{(i)}.
\end{eqnarray}
For a given $E$, there are two solutions $z_i$ ($z_1, z_2$), and thus they should fulfill the following constraint condition:
\begin{eqnarray}\label{XSz1z2}
z_{1}z_{2} &=&\frac{t_{1R}t_{2R}}{t_{1L}t_{2L}}.
\end{eqnarray}
Therefore, the superposition of two linearly independent solutions is also the solution of Eq.(\ref{XSbk1},\ref{XSbk2}) corresponding the same eigenvalue, i.e.,
\begin{equation}
\Psi=c_{1}\Psi _{1}+c_{2}\Psi _{2} = (\psi _{1A},\psi _{1B},\psi _{2A},\psi _{2B},\cdots ,\psi _{MA},\psi _{MB})^{T},~~~ \label{XSWave}
\end{equation}
where
\begin{equation}\label{XSWave1}
\psi _{nA}=\sum_{i=1}^{2}(c_{i}z_{i}^{n}\phi _{A}^{(i)})=c_{1}z_{1}^{n}\phi _{A}^{(1)}+c_{2}z_{2}^{n}\phi _{A}^{(2)},~~~\psi _{nB}=\sum_{i=1}^{2}(c_{i}z_{i}^{n}\phi _{B}^{(i)})=c_{1}z_{1}^{n}\phi _{B}^{(1)}+c_{2}z_{2}^{n}\phi _{B}^{(2)}
\end{equation}
with $n=1,2,\cdots ,M$.

\begin{figure*}[tbp]
\includegraphics[width=1.0\textwidth]{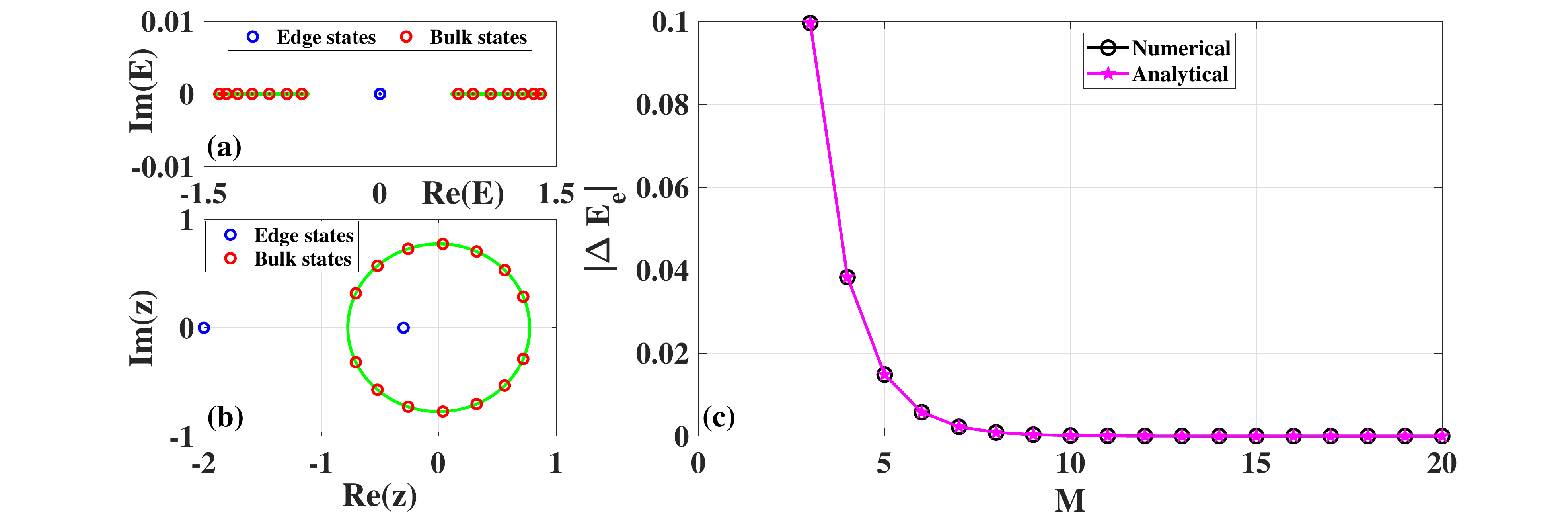}
\caption{(a, b) Energy spectra and the finite-size generalized Brillouin zones $z$ of non-Hermitian SSH model for $M=8$. The red circles and blue circles represent bulk states and edge states from analytical results, respectively. The red and blue dots in (a) represent the energy spectra from numerical results. The curve formed by green dots in (a) and (b) represents the energy spectra and the GBZ for OBC case in the thermodynamic limit, respectively. (c) The absolute value of energy splitting of two edge states as a function of cell size $M$ from numerical results (black circles) and analytical results (pink stars). Common parameters: $t_{1L}=0.5, t_{1R}=0.3, t_{2L}=t_{2R}=1$.}
\label{Xfig3}
\end{figure*}

To solve the eigen equation, the general ansatz of wave function Eq.(\ref{XSWave}) should satisfy the boundary conditions.
By inserting  the expression of $\Psi$ into Eqs.(\ref{XSbd3},\ref{XSbd4}),  the boundary equation transforms into
\begin{equation}\label{XSbb1}
H_{B}\left(
       \begin{array}{c}
         c_{1} \\
         c_{2} \\
       \end{array}
     \right)
=\left(
\begin{array}{cc}
(t_{2R}-\delta _{R}z_{1}^{M})\phi _{B}^{(1)} & (t_{2R}-\delta
_{R}z_{2}^{M})\phi _{B}^{(2)} \\
(t_{2L}z_{1}^{M}-\delta _{L})z_{1}\phi _{A}^{(1)} & (t_{2L}z_{2}^{M}-\delta
_{L})z_{2}\phi _{A}^{(2)}%
\end{array}%
\right)\left(
       \begin{array}{c}
         c_{1} \\
         c_{2} \\
       \end{array}
     \right)=0.
\end{equation}
The condition for the existence of nontrivial solutions for $(c_1, c_2)$ is determined by $\mathrm{det}[H_{B}] =0$,
which gives rise to the general solution:
\begin{equation}\label{XSqqz1}
(z_{1}^{M+1}-z_{2}^{M+1})+\chi_1(z_{1}^{M}-z_{2}^{M})-\chi_2(z_{1}^{M-1}-z_{2}^{M-1})
- \chi_3(z_{1}-z_{2})=0,
\end{equation}
where $\chi_1=\frac{t_{2R}t_{2L}-\delta _{R}\delta _{L}}{t_{1L}t_{2L}}$, $\chi_2=\frac{t_{1R}\delta _{R}\delta _{L}}{t_{1L}t_{2L}^{2}}$, $\chi_3= \frac{\delta _{L}}{t_{2L}}+\frac{\delta _{R}%
}{t_{2R}}\left( \frac{t_{1R}t_{2R}}{t_{1L}t_{2L}}\right) ^{M} $.
Eq.(\ref{XSqqz1}) and  Eq.(\ref{XSz1z2}) together determine the solution of $z_1$ and $z_2$ exactly.
According to the constraint condition of Eq.(\ref{XSz1z2}), we can always set the solution as
\begin{equation}
z_{1}=re^{i\theta}, ~~~ z_{2}=re^{-i\theta}
\end{equation}
with $r=\sqrt{\frac{t_{1R}t_{2R}}{t_{1L}t_{2L}}}$, then Eq.(\ref{XSqqz1})  becomes
\begin{equation}\label{XSeq-theta}
\sin[(M+1)\theta]+\eta_1\sin[M\theta]-\eta_2\sin[(M-1)\theta]-\eta_3\sin[\theta]=0,
\end{equation}
where $\eta_1=\frac{t_{2R}t_{2L}-\delta_R\delta_L}{\sqrt{t_{1R}t_{2R}t_{1L}t_{2L}}}$,   $\eta_2=\frac{\delta_R\delta_L}{t_{2R}t_{2L}}$ and $\eta_3=\frac{\delta _{L}}{t_{2L}} r ^{-M}+\frac{\delta _{R}}{t_{2R}} r ^{M}$.  The corresponding eigenvalue can be expressed as
\begin{equation}\label{XSspectrum1}
E=\pm\sqrt{2\sqrt{t_{1R}t_{2R}t_{1L}t_{2L}}\cos \theta+t_{1R}t_{1L}+t_{2R}t_{2L}} .
\end{equation}
The solution $\theta$ of Eq.(\ref{XSeq-theta}) may take real or complex depending on the values of $\eta_1$,  $\eta_2$ and $\eta_3$.

The OBC corresponds to the special case with $\delta _{R}=\delta _{L}=0$,
for which we have $\eta_1=\alpha=\sqrt{\frac{t_{2R}t_{2L}}{t_{1R}t_{1L}}}$, $\eta_2=\eta_3=0$, and Eq.(\ref{XSeq-theta}) can be rewritten as
\begin{equation}\label{XSeq-theta1}
\sin[(M+1)\theta]+\alpha\sin[M\theta]=0.
\end{equation}
We can see that Eq.(\ref{XSeq-theta1}) has $M$ real solutions corresponding bulk states when
$\alpha<\alpha_c$, while $M-1$ real solutions corresponding bulk states and 1 complex solution ($\theta=\pi+i\varphi$) corresponding edge states when
$\alpha>\alpha_c$. Here we set $f_1=\sin[(M+1)\theta], f_2=-\alpha\sin[M\theta]$, and $\alpha_c$ is determined by $f_1'(\theta=\pi)=f_2'(\theta=\pi)$, where $f_{i}'=\frac{\partial f_i}{\partial \theta}$. It follows
\begin{equation}
\alpha_c=1+\frac{1}{M}.
\end{equation}
In the thermodynamic limit, we have $\alpha_c=1$, and thus the boundary of topological phase transition is given by $\alpha=1$, i.e., $|t_{2R}t_{2L}|=|t_{1R}t_{1L}|$. In the topological phase ($\alpha>\alpha_c$), the only complex solution of $\theta$ is $\theta=\pi+i\varphi, (\varphi\in R)$,
and Eq.(\ref{XSeq-theta}) becomes
\begin{equation}\label{XSeq-theta2}
\sinh[(M+1)\varphi]-\alpha\sinh[M\varphi]=0.
\end{equation}
Due to $\sinh[x]=(e^x-e^{-x})/2$, the above equation is equivalent to
\begin{equation}
2M\varphi =\log (\frac{e^{-\varphi }-\alpha }{e^{\varphi }-\alpha }).
\end{equation}
This equation has solution only when $e^{\varphi }\approx \alpha $. If we set $e^{\varphi
}=\alpha +\delta_\varphi (\delta_\varphi \rightarrow 0)$, then we have
\begin{equation}
\delta_\varphi (\alpha +\delta_\varphi )^{2M} =\frac{1}{\alpha +\delta_\varphi }-\alpha.
\end{equation}
Due to $\delta_\varphi \rightarrow 0$, we have $(\alpha +\delta_\varphi)^{2M}\approx\alpha^{2M}$, $\frac{1}{\alpha +\delta_\varphi }\approx \frac{1}{\alpha }-\frac{\delta_\varphi }{\alpha ^{2}}$, then we obtain
\begin{eqnarray}
\delta_\varphi =\frac{(\frac{1}{\alpha }-\alpha )}{(\alpha ^{2M}+\frac{1}{\alpha^2})}.
\end{eqnarray}
Therefore, we have
\begin{eqnarray}
\varphi=\log \left[ \alpha \left( 1+\frac{%
1-\alpha ^{2}}{1+\alpha ^{2(M+1)}}%
\right) \right]
\end{eqnarray}
which is dependent on $M$, and the energy corresponding edge states becomes
\begin{equation}\label{XSEz1}
E_e= \pm \sqrt{-2\sqrt{t_{1R}t_{2R}t_{1L}t_{2L}}\cosh(\varphi)
+t_{1L}t_{1R}+t_{2L}t_{2R}} .
\end{equation}
In the thermodynamic limit,  $\varphi \rightarrow \log (\alpha)$ and $E_e\rightarrow 0$, indicating the emergence of degenerate zero modes.

In Figs.\ref{Xfig3}(a, b), we present energy spectra and the finite-size generalized Brillouin zones $z$ of non-Hermitian SSH model for $M=8$. In Fig.\ref{Xfig3}(c), we plot the absolute value of energy splitting of two edge states ($|\Delta E_e|=2|E_e|$) as a function of cell size $M$ from numerical results (black circles) and analytical results (pink stars). The numerical results are consistent with our analytical predictions.

Then we calculate the eigenstates for OBC case, from Eq.(\ref{XSbb1}), we can obtain the following relation for bulk states as
\begin{eqnarray}\label{XSbab}
c_{2}\phi _{A}^{(2)}=-\frac{(t_{1R}+t_{2L}z_{1})}{(t_{1R}+t_{2L}z_{2})}%
c_{1}\phi _{A}^{(1)}=-\frac{1+\alpha e^{i\theta }}{1+\alpha e^{-i\theta }}%
c_{1}\phi _{A}^{(1)}.
\end{eqnarray}
By inserting Eq.(\ref{XSbab}) into Eq.(\ref{XSWave1}), we can obtain $\psi _{nA/B}$ for bulk states as
\begin{eqnarray}
\psi _{nA} &=&\sum_{i=1}^{2}(c_{i}z_{i}^{n}\phi
_{A}^{(i)})=c_{1}z_{1}^{n}\phi _{A}^{(1)}+c_{2}z_{2}^{n}\phi _{A}^{(2)}\text{
\ \ \ \ }(n=1,2,\cdots ,M) \nonumber\\
&=&r^{n}e^{in\theta }c_{1}\phi _{A}^{(1)}+r^{n}e^{-in\theta }c_{2}\phi
_{A}^{(2)}\text{ }  \nonumber\\
&=&\frac{2ic_{1}\phi _{A}^{(1)}}{1+\alpha e^{-i\theta }}r^{n}\left( \sin
[n\theta ]+\alpha \sin [(n-1)\theta ]\right)   \nonumber\\
&\sim &r^{n}\left( \sin [n\theta ]+\alpha \sin [(n-1)\theta ]\right),
\end{eqnarray}
\begin{eqnarray}
\psi _{nB} &=&\sum_{i=1}^{2}(c_{i}z_{i}^{n}\phi
_{B}^{(i)})=c_{1}z_{1}^{n}\phi _{B}^{(1)}+c_{2}z_{2}^{n}\phi _{B}^{(2)}\text{
\ \ \ \ }(n=1,2,\cdots ,M)  \nonumber\\
&=&r^{n}e^{in\theta }c_{1}\phi _{B}^{(1)}+r^{n}e^{-in\theta }c_{2}\phi
_{B}^{(2)}\text{ }  \nonumber\\
&=&\frac{2ic_{1}\phi _{A}^{(1)}}{1+\alpha e^{-i\theta }}\frac{E}{t_{1L}}%
r^{n}\sin [n\theta ]  \nonumber\\
&\sim &\frac{E}{t_{1L}}r^{n}\sin [n\theta ],
\end{eqnarray}
where $r=\sqrt{\frac{t_{1R}t_{2R}}{t_{1L}t_{2L}}}, \alpha =\sqrt{%
\frac{t_{2R}t_{2L}}{t_{1R}t_{1L}}}$, $\theta$ is real number which satisfies Eq.(\ref{XSeq-theta1}), and $E$ is the corresponding energy which satisfies Eq.(\ref{XSspectrum1}). From the eigenstates for bulk states, we can see that there is skin effect for OBC case.

Similarly, from Eq.(\ref{XSbb1}), we can obtain the following relation for edge states as
\begin{eqnarray}\label{XSeab}
c_{2}\phi _{A}^{(2)}=-\frac{(t_{1R}+t_{2L}z_{1})}{(t_{1R}+t_{2L}z_{2})}%
c_{1}\phi _{A}^{(1)}=-\frac{1-\alpha e^{\varphi }}{1-\alpha e^{-\varphi }}%
c_{1}\phi _{A}^{(1)}\text{ \ \ \ \ }(\alpha =\sqrt{\frac{t_{2R}t_{2L}}{%
t_{1R}t_{1L}}}),
\end{eqnarray}
By inserting Eq.(\ref{XSeab}) into Eq.(\ref{XSWave1}), we can obtain $\psi _{nA/B}$ for edge states as
\begin{eqnarray}
\psi _{nA} &=&\sum_{i=1}^{2}(c_{i}z_{i}^{n}\phi
_{A}^{(i)})=c_{1}z_{1}^{n}\phi _{A}^{(1)}+c_{2}z_{2}^{n}\phi _{A}^{(2)}\text{
\ \ \ \ }(n=1,2,\cdots ,M)  \nonumber\\
&=&(-1)^{n}r^{n}e^{n\varphi }c_{1}\phi _{A}^{(1)}+(-1)^{n}r^{n}e^{-n\varphi }c_{2}\phi
_{A}^{(2)}\text{ }  \nonumber\\
&=&(-1)^{n}\frac{2ic_{1}\phi _{A}^{(1)}}{1-\alpha e^{-\varphi }}r^{n}\left( \sinh
[n\varphi]-\alpha \sinh [(n-1)\varphi ]\right)   \nonumber\\
&\sim &(-1)^{n}r^{n}\left( \sinh [n\varphi]-\alpha \sinh [(n-1)\varphi]\right),
\end{eqnarray}
\begin{eqnarray}
\psi _{nB} &=&\sum_{i=1}^{2}(c_{i}z_{i}^{n}\phi
_{B}^{(i)})=c_{1}z_{1}^{n}\phi _{B}^{(1)}+c_{2}z_{2}^{n}\phi _{B}^{(2)}\text{
\ \ \ \ }(n=1,2,\cdots ,M)  \nonumber\\
&=&(-1)^{n}r^{n}e^{n\varphi }c_{1}\phi _{B}^{(1)}+(-1)^{n}r^{n}e^{-n\varphi }c_{2}\phi
_{B}^{(2)}\text{ }  \nonumber\\
&=&(-1)^{n}\frac{2ic_{1}\phi _{A}^{(1)}}{1-\alpha e^{-\varphi }}\frac{E}{t_{1L}}%
r^{n}\sinh [n\varphi ]  \nonumber\\
&\sim &(-1)^{n}\frac{E}{t_{1L}}r^{n}\sinh [n\varphi ],
\end{eqnarray}
where $r=\sqrt{\frac{t_{1R}t_{2R}}{t_{1L}t_{2L}}}, \alpha =\sqrt{%
\frac{t_{2R}t_{2L}}{t_{1R}t_{1L}}}$, $\varphi$ is given by Eq.(\ref{XSeq-theta2}), and $E$ is the corresponding energy which is given by Eq.(\ref{XSEz1}).

For cases with either $\delta _{R}=0$ ($\delta _{L}\neq0$) or $\delta _{L}=0$ ($\delta _{R} \neq 0$), we have $\eta_1=\alpha$, $\eta_2=0$, and $\eta_3=\frac{\delta _{L}}{t_{2L}} r ^{-M}$ or $\eta_3=\frac{\delta _{R}}{t_{2R}} r ^{M}$. In the thermodynamic limit, we have $\eta_3\rightarrow 0$ for the case of $\delta_R=0$ and $r>1$ ($t_{1R}/t_{2L}>t_{1L}/t_{2R}$) or $\delta_L=0$ and $r<1$ ($t_{1R}/t_{2L}<t_{1L}/t_{2R}$), and the solutions of Eq.(\ref{XSeq-theta}) are identical to the OBC case. The analytical results indicate clearly that in these cases the system exhibits skin effect as all wavefunctions accumulated either on the left ($r<1$) or right ($r>1$) edge in the large size limit.

As a contrast, for the cases of $\delta _{R}=0$ ($\delta _{L}\neq0$) with $r<1$, the case of $\delta _{L}=0$ ($\delta _{R} \neq 0$) with $r>1$, and the case of $\delta _{R}\neq0, \delta _{L}\neq0$, we have $\eta_3\rightarrow\infty$ in the thermodynamic limit. In the region of $0\leq\frac{\delta_R}{t_R}, \frac{\delta_L}{t_L}\leq1$, we have $\eta_1\geq 0$, $0\leq\eta_2\leq 1$ and $\eta_3\geq 0$. Eq.(\ref{XSeq-theta}) can be rewritten as
\begin{equation}\label{XSeq-thetaa}
f_1=f_2
\end{equation}
with
\begin{equation}
f_1=(1+\eta_2)\cos[M\theta]\sin[\theta]+\left(\eta_1+(1-\eta_2)\cos[\theta]\right)\sin[M\theta], ~~~~~~f_2=\eta_3\sin[\theta],
\end{equation}
where $\eta_1=\frac{t_{2R}t_{2L}-\delta_R\delta_L}{\sqrt{t_{1R}t_{2R}t_{1L}t_{2L}}}$,   $\eta_2=\frac{\delta_R\delta_L}{t_{2R}t_{2L}}$ and $\eta_3=\frac{\delta _{L}}{t_{2L}} r ^{-M}+\frac{\delta _{R}}{t_{2R}} r ^{M}$. For $\theta\in(0,\pi)$, the outer contour of $f_1$ is larger than $\max\{ \pm(1+\eta_2)\sin[\theta], \pm\left(\eta_1+(1-\eta_2)\cos[\theta]\right)\}$. Because the last real solution for  Eq.(\ref{XSeq-thetaa}) to disappear is near $\theta=0$, the condition with no real solution but complex solutions is $f_2'(\theta=0)>f_1'(\theta=0)$, i.e. $\eta_3>M+1+\eta_1M-\eta_2(M-1)$, which is easily satisfied for these cases with $\eta_3\rightarrow\infty$ in the thermodynamic limit.

\section{Fragility of non-Hermitian skin effect for 2D models with generalized boundary conditions}

We have shown that the skin effect in the one-dimensional nonreciprocal lattices is fragile under a tiny boundary perturbation in the thermodynamic limit. Such a phenomenon is expected to be observed in higher-dimensional systems. To see it clearly, we first consider a concrete example, i.e.,  a 2D generalization of Hatano-Nelson model, which exhibits non-Hermitian skin effect under OBC as demonstrated in Ref.\cite{GongJBSM}. To understand the fate of non-Hermitian skin effect under the boundary perturbations, we consider the 2D skin model with generalized boundary conditions, which is described by
\begin{equation}
\begin{split}
\hat{H}_{\mathrm{skin}}^{\mathrm{2D}}=&\sum\limits_{i=1}^{N_x-1}\sum\limits_{j=1}^{N_y}\left[ t_{L}^{x}\hat{c}_{i,j}^{\dag
}\hat{c}_{i+1,j}+t_{R}^{x}\hat{c}_{i+1,j}^{\dag
}\hat{c}_{i,j}\right] +\sum\limits_{i=1}^{N_x}\sum\limits_{j=1}^{N_y-1}\left[ t_{L}^{y}\hat{c}_{i,j}^{\dag
}\hat{c}_{i,j+1}+t_{R}^{y}\hat{c}_{i,j+1}^{\dag
}\hat{c}_{i,j}\right]\\
&+\sum\limits_{j=1}^{N_y}\left[ \delta_{L}^{x}\hat{c}_{N_x,j}^{\dag
}\hat{c}_{1,j}+\delta_{R}^{x}\hat{c}_{1,j}^{\dag
}\hat{c}_{N_x,j}\right]+\sum\limits_{i=1}^{N_x}\left[\delta_{L}^{y}\hat{c}_{i,N_y}^{\dag
}\hat{c}_{i,1}+\delta_{R}^{y}\hat{c}_{i,1}^{\dag
}\hat{c}_{i,N_y}\right],
\end{split}
\end{equation}
where $t_{L}^{x}, t_{R}^{x}, t_{L}^{y}, t_{R}^{y}\in \mathbb{R}$ are imbalanced hopping amplitudes, $\delta_{L}^{x}, \delta_{R}^{x}, \delta_{L}^{y}, \delta_{R}^{y} \in \mathbb{R}$ determine the generalized boundary conditions, and $N_x/N_y$ is the number of lattice sites along $x/y$ direction. Similarly, the corresponding eigenvalue equation can be written as
$
\hat{H}_{\mathrm{skin}}^{\mathrm{2D}} |\Psi\rangle = E|\Psi\rangle
$,
where $|\Psi\rangle=\sum_{x,y}\psi_{x,y}|x,y\rangle= \sum_{x,y}\psi_{x,y} \hat{c}_{x,y}^{\dag
} |0\rangle$.

In the following, we discuss the fragility of non-Hermitian skin effect in two cases:
(1) We apply PBC in the y-direction and generalized boundary condition in the x-direction;
(2) We apply generalized boundary condition in both the x-direction and y-direction.

Firstly, when we apply PBC in the y-direction (i.e. $\delta_{L}^{y}=t_L^y, \delta_{R}^{y}=t_R^y$), we can perform Fourier transformation for the y-direction
\begin{equation}
\hat{c}_{i,j} = \frac{1}{\sqrt{N_y}}\sum_{k_y}  e^{i k_y j} \hat{c}_{i,k_y},~~~~
\hat{c}_{i,j}^{\dag} = \frac{1}{\sqrt{N_y}}\sum_{k_y} e^{-i k_y j} \hat{c}_{i,k_y}^{\dag}.
\end{equation}
Then the Hamiltonian becomes $\hat{H}_{\mathrm{skin}}^{\mathrm{2D}}=\sum_{k_y}\hat{H}_{\mathrm{skin}}^{\mathrm{2D}}(k_y)$ with
\begin{equation}
\hat{H}_{\mathrm{skin}}^{\mathrm{2D}}(k_y)=\sum\limits_{i=1}^{N_x-1}\left[ t_{L}^{x}\hat{c}_{i,k_y}^{\dag
}\hat{c}_{i+1,k_y}+t_{R}^{x}\hat{c}_{i+1,k_y}^{\dag
}\hat{c}_{i,k_y}\right]+\left[ \delta_{L}^{x}\hat{c}_{N_x,k_y}^{\dag
}\hat{c}_{1,k_y}+\delta_{R}^{x}\hat{c}_{1,k_y}^{\dag
}\hat{c}_{N_x,k_y}\right] +\left(t_{L}^{y}e^{ik_y}+t_{R}^{y}e^{-ik_y} \right)\sum\limits_{i=1}^{N_x}\hat{c}_{i,k_y}^{\dag
}\hat{c}_{i,k_y}.
\end{equation}
The Hamiltonian above $\hat{H}_{\mathrm{skin}}^{\mathrm{2D}}(k_y)$ is similar to the Hamiltonian of Hatano-Nelson model of Eq.(1) in the main text by performing replacements $t_{L}^{x} \rightarrow t_{L}, t_{R}^{x} \rightarrow t_{R}, \delta_{L}^{x},  \rightarrow \delta_{L},\delta_{R}^{x} \rightarrow \delta_{R}, N_x  \rightarrow N, \hat{c}_{i,k_y}^{\dag}\rightarrow \hat{c}_{i}^{\dag}, \hat{c}_{i,k_y}\rightarrow \hat{c}_{i}$ with an additional on-site terms.
\begin{figure*}[tbp]
\includegraphics[width=1.0\textwidth]{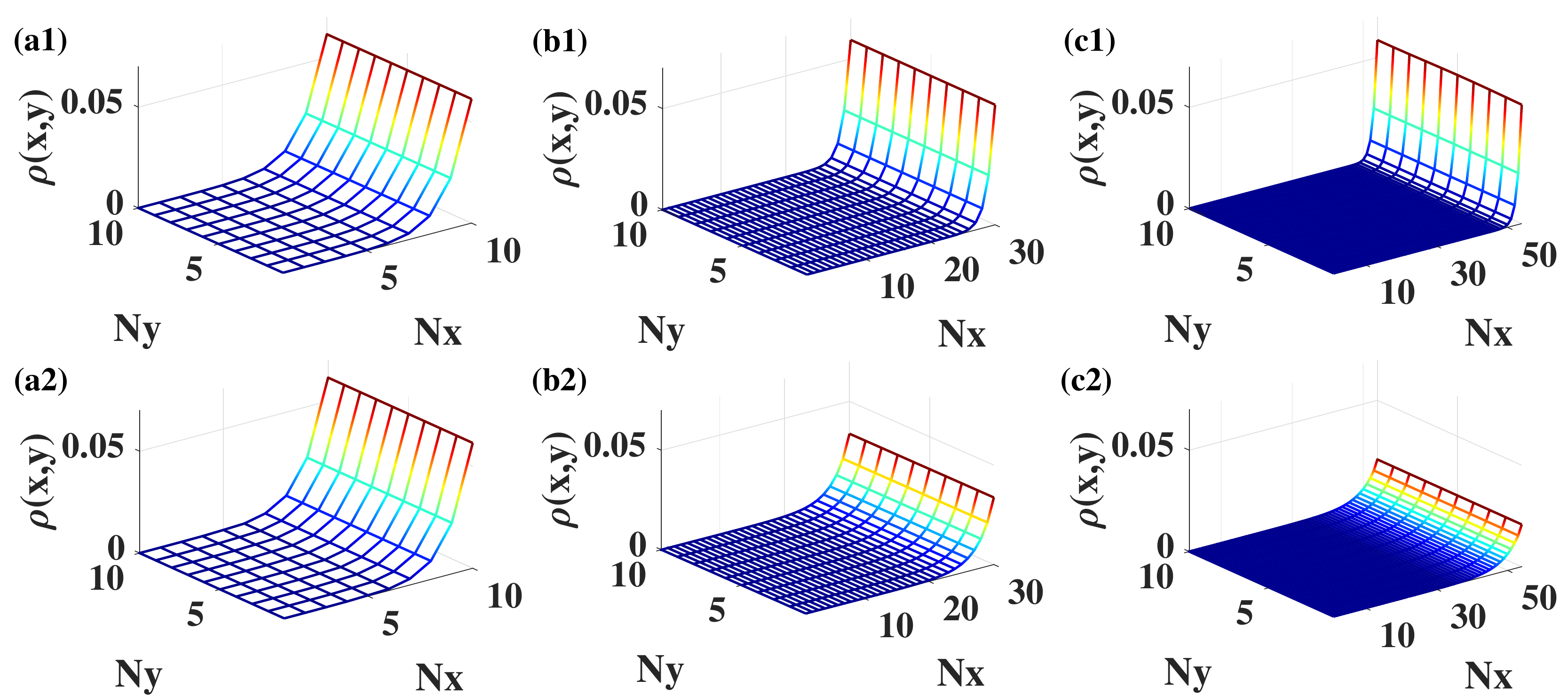}
\caption{(a1,b1,c1) The profile of eigenstates $\rho(x,y)$ for $\hat{H}_{\mathrm{skin}}^{\mathrm{2D}}$ with OBC in the x-direction and PBC in the y-direction for $N_x*N_y=10*10, 30* 10, 54*10$, respectively.
(a2,b2,c2) The profile of eigenstates $\rho(x,y)$ for $\hat{H}_{\mathrm{skin}}^{\mathrm{2D}}$ with generalized boundary condition in the x-direction ($\delta_{R}^x=\delta_{L}^x=0.006$) and PBC in the y-direction for $N_x*N_y=10*10, 30* 10, 54*10$, respectively. Common parameters: $t_{R}^x=t_{R}^y=2.3, t_{L}^x=t_{L}^y=0.7, \delta_{R}^y=t_{R}^y, \delta_{L}^y=t_{L}^y$.}
\label{2Dfig1}
\end{figure*}
\begin{figure*}[tbp]
\includegraphics[width=1.0\textwidth]{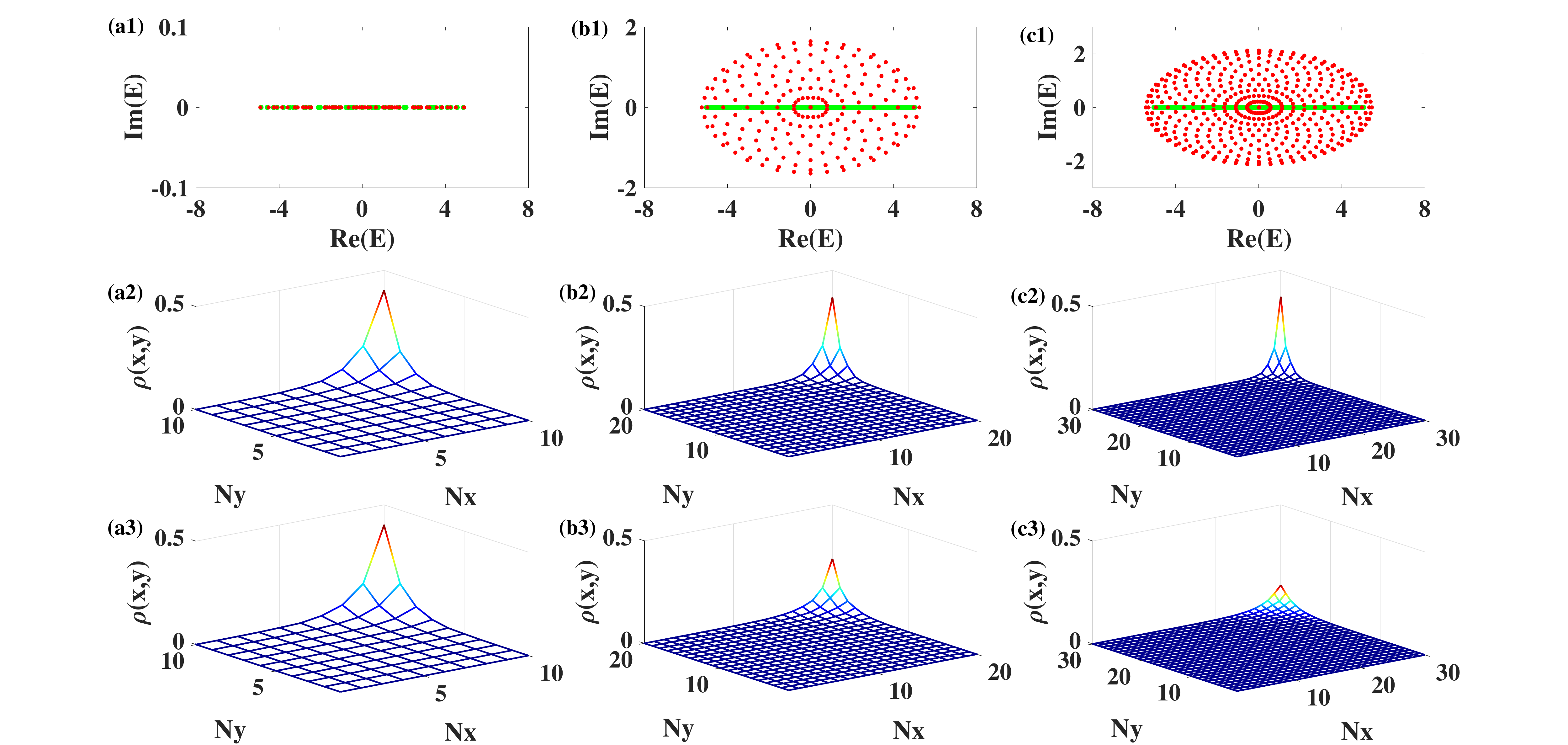}
\caption{(a1,b1,c1) Energy spectra of $\hat{H}_{\mathrm{skin}}^{\mathrm{2D}}$ for $N_x*N_y=10*10, 20* 20, 30*30$, respectively. The green dots represent the energy spectra with OBC in both the x-direction and y-direction, and the red dots represent the energy spectra with generalized boundary condition in the x-direction and y-direction ($\delta_{R}^x=\delta_{L}^x=\delta_{R}^y=\delta_{L}^y=0.006$).
(a2,b2,c2) The profile of eigenstates $\rho(x,y)$ of $\hat{H}_{\mathrm{skin}}^{\mathrm{2D}}$ with OBC in both the x-direction and y-direction for $N_x*N_y=10*10, 20* 20, 30*30$, respectively.
(a3,b3,c3) The profile of eigenstates $\rho(x,y)$ of $\hat{H}_{\mathrm{skin}}^{\mathrm{2D}}$ with generalized boundary condition in the x-direction and y-direction ($\delta_{R}^x=\delta_{L}^x=\delta_{R}^y=\delta_{L}^y=0.006$) for $N_x*N_y=10*10, 20* 20, 30*30$, respectively. Common parameters: $t_{R}^x=t_{R}^y=2.3, t_{L}^x=t_{L}^y=0.7$.}
\label{2Dfig2}
\end{figure*}

Similarly, the corresponding eigenvalue equation can be written as
$
\hat{H}_{\mathrm{skin}}^{\mathrm{2D}}(k_y)|\Psi(k_y)\rangle = E(k_y)|\Psi(k_y)\rangle
$,
where $|\Psi(k_y)\rangle=\sum_{x}\psi_{x}(k_y)|x\rangle = \sum_{x}\psi_{x}(k_y) \hat{c}_{x,k_y}^{\dag
} | 0 \rangle  $. For convenience, we also denote $\Psi(k_y) = (\psi_{1},\psi_{2},\cdots,\psi_{N_x})^T$, here we simplify $\psi_{x}(k_y)$ to $\psi_{x}$.
The above eigenvalue equation consists of a series of equations, including bulk equations as follows
\begin{equation}\label{2Dbk}
t_{R}^x\psi _{s}-\left[E(k_y)-t_{L}^{y}e^{ik_y}-t_{R}^{y}e^{-ik_y} \right]\psi _{s+1}+t_{L}^x\psi _{s+2}=0
\end{equation}%
with $s=1,2,\cdots,N_x-2$, and the boundary equations given by
$-\left[E(k_y)-t_{L}^{y}e^{ik_y}-t_{R}^{y}e^{-ik_y} \right]\psi _{1}+t_{L}^x\psi _{2}+\delta _{R}^x\psi _{N_x} =  0$ and
$\delta _{L}^x\psi _{1}+t_{R}^x\psi _{N-1}-\left[E(k_y)-t_{L}^{y}e^{ik_y}-t_{R}^{y}e^{-ik_y} \right]\psi _{N_x} =  0$.
By comparing the above two equations with Eq.(\ref{2Dbk}), they are equivalent to the following boundary conditions
\begin{eqnarray}
t_{R}^x\psi _{0} &=&\delta _{R}^x\psi _{N_x},  \label{2Dbd3} \\
\delta _{L}^x\psi _{1} &=&t_{L}^x\psi _{N_x+1}. \label{2Dbd4}
\end{eqnarray}
Due to spatial translational property from bulk equations, we set the ansatz of wave function $\Psi _{i}(k_y)$ which satisfies the bulk equations Eq.(\ref{2Dbk}) as follows
\begin{equation}\label{2DFii}
\Psi _{i}(k_y)=(z_{i},z_{i}^{2},z_{i}^{3},\cdots ,z_{i}^{N_x-1},z_{i}^{N_x})^{T}.
\end{equation}
By inserting  Eq.(\ref{2DFii}) into the bulk equation Eq.(\ref{2Dbk}), we obtain the expression of eigenvalue in terms of $z_i$ as
\begin{eqnarray}\label{2DEz1}
E(k_y) &=&\frac{t_{R}^x}{z_{i}}+t_{L}^xz_{i}+t_{L}^{y}e^{ik_y}+t_{R}^{y}e^{-ik_y}  .
\end{eqnarray}
For a given $E(k_y)$, there are two solutions $z_i$ ($z_1, z_2$), and thus they should fulfill the following constraint condition:
\begin{equation}\label{2Dz1z2}
z_{1}z_{2}=\frac{t_{R}^x}{t_{L}^x} .
\end{equation}
Therefore, the superposition of two linearly independent solutions is also the solution of Eq.(\ref{2Dbk}) corresponding the same eigenvalue $E(k_y)$, i.e.,
\begin{equation}
\Psi(k_y)=c_{1}\Psi _{1}(k_y)+c_{2}\Psi _{2}(k_y) = (\psi _{1},\psi _{2},\cdots ,\psi _{N_x})^{T}~~~ \label{2DWave}
\end{equation}
where $\psi _{n}=\sum_{i=1}^{2}(c_{i}z_{i}^{n})=c_{1}z_{1}^{n}+c_{2}z_{2}^{n}$
with $n=1,2,\cdots ,N_x$.

The solutions of $z_1, z_2$ can be obtained by inserting the expression of $\Psi(k_y)$ into boundary conditions Eq.(\ref{2Dbd3}) and Eq.(\ref{2Dbd4}). Because Eqs.(\ref{2Dbd3}, \ref{2Dbd4}) are equivalent to Eq.(3) and Eq.(4) in the main text by performing replacements $t_{L/R}^{x}\rightarrow t_{L}, \delta_{L/R}^{x}\rightarrow \delta_{L/R}, N_x \rightarrow N$, the solutions of $z_1, z_2$ for this case are the same as those for Hatano-Nelson model. Therefore, the wave function $\Psi(k_y)$ along x-direction exhibts similar behavior as the eigenfunction $\Psi$ of Hatano-Nelson model.

The amplitude of non-Hermitian skin effect can be quantified by the averaged squared eigenmode amplitude defined by
\begin{equation}
\rho(x,y)=\frac{1}{N_xN_y}\sum_{s}|\langle x,y|\Psi^{s}\rangle|^2,
\end{equation}
where $|\Psi^{s}\rangle$ is the s-th eigenfunctions $|\Psi\rangle$ of $\hat{H}_{\mathrm{skin}}^{\mathrm{2D}}$ and the summation runs over all eigenfunctions.

When we apply OBC in the x-direction, the system exhibits non-Hermitian skin effect along x-direction as all wave functions accumulate on the edge independent of lattice size $N_x$, as plotted in Figs.\ref{2Dfig1}(a1,b1,c1). However, the NHSE along x-direction under tiny boundary perturbations along x-direction is fragile in the thermodynamic limit. As shown in Figs.\ref{2Dfig1}(a2,b2,c2), the non-Hermitian skin effect along x-direction is diminished as the lattice size $N_x$ increases.

Next, we  consider  the case with generalized boundary condition in both the x-direction and y-direction, which can not be analytically solved. We numerically diagonalize the finite-size systems and demonstrate the fragility of the non-Hermitian skin effect under tiny boundary perturbations in the large size limit. In Figs.\ref{2Dfig2}(a1,b1,c1), we display the energy spectra for $\hat{H}_{\mathrm{skin}}^{\mathrm{2D}}$ with different sizes under both OBC and generalized boundary condition. While the spectra for different size systems under OBC are always real, complex spectrum for the system under generalized boundary condition emerges when the system size increases. Our results unveil that the spectrum is sensitive to the boundary perturbation, which suggests the fragility of non-Hermitian skin effect under tiny boundary perturbations in the large size limit. To see it more clearly, we display profiles of eigenstates $\rho(x,y)$ for $\hat{H}_{\mathrm{skin}}^{\mathrm{2D}}$ under OBC in Figs.\ref{2Dfig2}(a2,b2,c2) and the generalized boundary condition  in Figs.\ref{2Dfig2}(a3,b3,c3), respectively.  It is shown that the system under OBC exhibits non-Hermitian skin effect along x-direction and y-direction as all wave functions accumulate on the corner independent of lattice size $N_x \times N_y$\cite{GongJBSM}, as plotted in Figs.\ref{2Dfig2}(a2,b2,c2). The profiles of eigenstates $\rho(x,y)$ under tiny boundary perturbations is similar to that under OBC for small lattice size as plotted in Figs.\ref{2Dfig2}(a2) and (a3). As the lattice size $N_x \times N_y$ increases,  the non-Hermitian skin effect is diminished under tiny boundary perturbations  as shown in Figs.\ref{2Dfig2}(b3) and (c3). Both the spectra and profiles of eigensates indicate that the non-Hermitian skin effect under tiny boundary perturbations is fragile in the thermodynamic limit for $\hat{H}_{\mathrm{skin}}^{\mathrm{2D}}$. 

Moreover, we consider another concrete example, i.e.,  a 2D second-order topological insulator (SOTI), which exhibits non-Hermitian skin effect under OBC as demonstrated in Ref.\cite{TLiuSM}.  The model with generalized boundary conditions is described by
\begin{equation}
\begin{split}
\hat{H}_{\mathrm{SOTI}}^{\mathrm{2D}}=&\sum\limits_{i=1}^{M_x}\sum\limits_{j=1}^{M_y}\big\{(t+\gamma)[\hat{c}_{i,j,C}^{\dag
}\hat{c}_{i,j,A}+\hat{c}_{i,j,B}^{\dag}\hat{c}_{i,j,C}-\hat{c}_{i,j,D}^{\dag}\hat{c}_{i,j,A}+\hat{c}_{i,j,B}^{\dag}\hat{c}_{i,j,D}]\\
&+(t-\gamma)[\hat{c}_{i,j,A}^{\dag}\hat{c}_{i,j,C}+\hat{c}_{i,j,C}^{\dag}\hat{c}_{i,j,B}-\hat{c}_{i,j,A}^{\dag
}\hat{c}_{i,j,D}+\hat{c}_{i,j,D}^{\dag}\hat{c}_{i,j,B}]\big\}\\
&+\sum\limits_{i=1}^{M_x-1}\sum\limits_{j=1}^{M_y}\lambda[\hat{c}_{i,j,A}^{\dag}\hat{c}_{i+1,j,C}+
\hat{c}_{i,j,D}^{\dag}\hat{c}_{i+1,j,B}+\hat{c}_{i+1,j,C}^{\dag}\hat{c}_{i,j,A}+\hat{c}_{i+1,j,B}^{\dag}\hat{c}_{i,j,D}]\\
&+\sum\limits_{i=1}^{M_x}\sum\limits_{j=1}^{M_y-1}\lambda[\hat{c}_{i,j,C}^{\dag}\hat{c}_{i,j+1,B}
-\hat{c}_{i,j,A}^{\dag}\hat{c}_{i,j+1,D}+\hat{c}_{i,j+1,B}^{\dag}\hat{c}_{i,j,C}-\hat{c}_{i,j+1,D}^{\dag}\hat{c}_{i,j,A}]\\
&+\sum\limits_{j=1}^{M_y}\big\{\delta_L^x[\hat{c}_{M_x,j,A}^{\dag}\hat{c}_{1,j,C}+
\hat{c}_{M_x,j,D}^{\dag}\hat{c}_{1,j,B}]+\delta_R^x[\hat{c}_{1,j,C}^{\dag}\hat{c}_{M_x,j,A}+\hat{c}_{1,j,B}^{\dag}\hat{c}_{M_x,j,D}]\big\}\\
&+\sum\limits_{i=1}^{M_x}\big\{\delta_L^y[\hat{c}_{i,M_y,C}^{\dag}\hat{c}_{i,1,B}-\hat{c}_{i,M_y,A}^{\dag}\hat{c}_{i,1,D}]+\delta_R^y
[\hat{c}_{i,1,B}^{\dag}\hat{c}_{i,M_y,C}-\hat{c}_{i,1,D}^{\dag}\hat{c}_{i,M_y,A}]\big\},\\
\end{split}
\end{equation}
where $t\pm\gamma\in \mathbb{R}$ are imbalanced intracell hopping amplitudes, $\lambda\in \mathbb{R}$ is a intercell hopping amplitude, $\delta_{L}^{x}, \delta_{R}^{x}, \delta_{L}^{y}, \delta_{R}^{y} \in \mathbb{R}$ determine the generalized boundary conditions, and $M_x/M_y$ is the number of unit cells along $x/y$ direction. Each unit cell
contains four sublattice labeled as A, B, C, D. Similarly, the corresponding eigenvalue equation can be written as
$
\hat{H}_{\mathrm{SOTI}}^{\mathrm{2D}} |\Psi\rangle = E|\Psi\rangle
$,
where $|\Psi\rangle=\sum_{x,y,n}\psi_{x,y,n}|x,y,n\rangle= \sum_{x,y,n}\psi_{x,y,n} \hat{c}_{x,y,n}^{\dag} |0\rangle$ with $n=A, B, C, D$.
\begin{figure*}[tbp]
\includegraphics[width=1.0\textwidth]{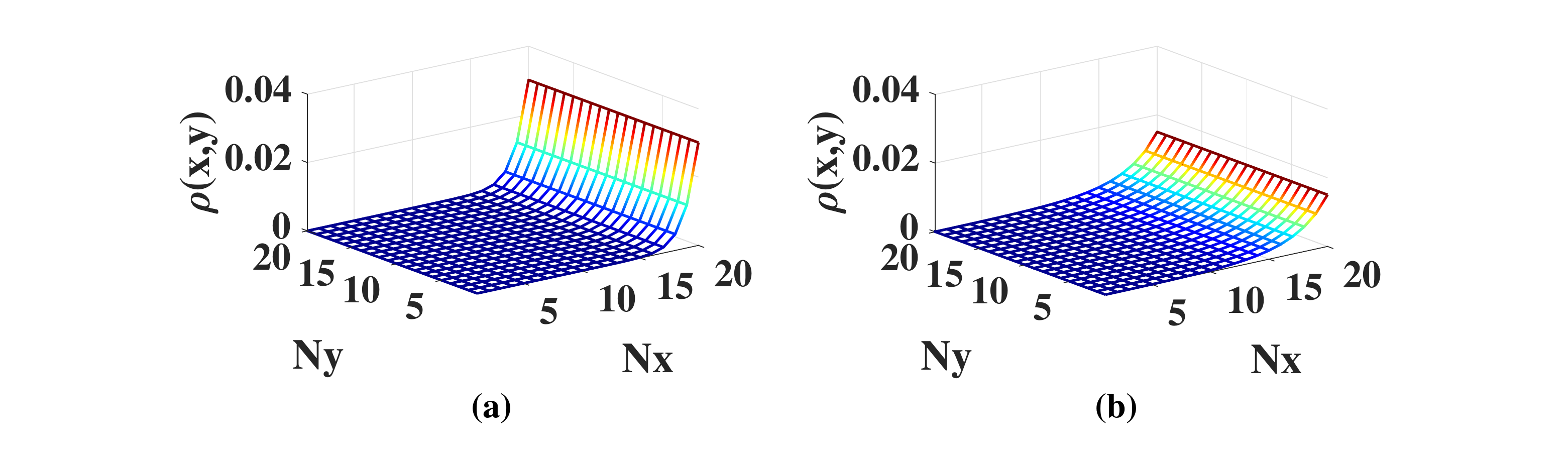}
\caption{(a) The profile of eigenstates $\rho(x,y)$ for $\hat{H}_{\mathrm{SOTI}}^{\mathrm{2D}}$ with OBC in the x-direction and PBC in the y-direction. (b) The profile of eigenstates $\rho(x,y)$ for $\hat{H}_{\mathrm{SOTI}}^{\mathrm{2D}}$ with generalized boundary condition in the x-direction ($\delta_{R}^x=\delta_{L}^x=0.02$) and PBC in the y-direction. Common parameters: $t=1.5, \gamma=-0.8, \lambda=0.6, M_x*M_y=20*20$.}
\label{SOTI1}
\end{figure*}
\begin{figure*}[tbp]
\includegraphics[width=1.0\textwidth]{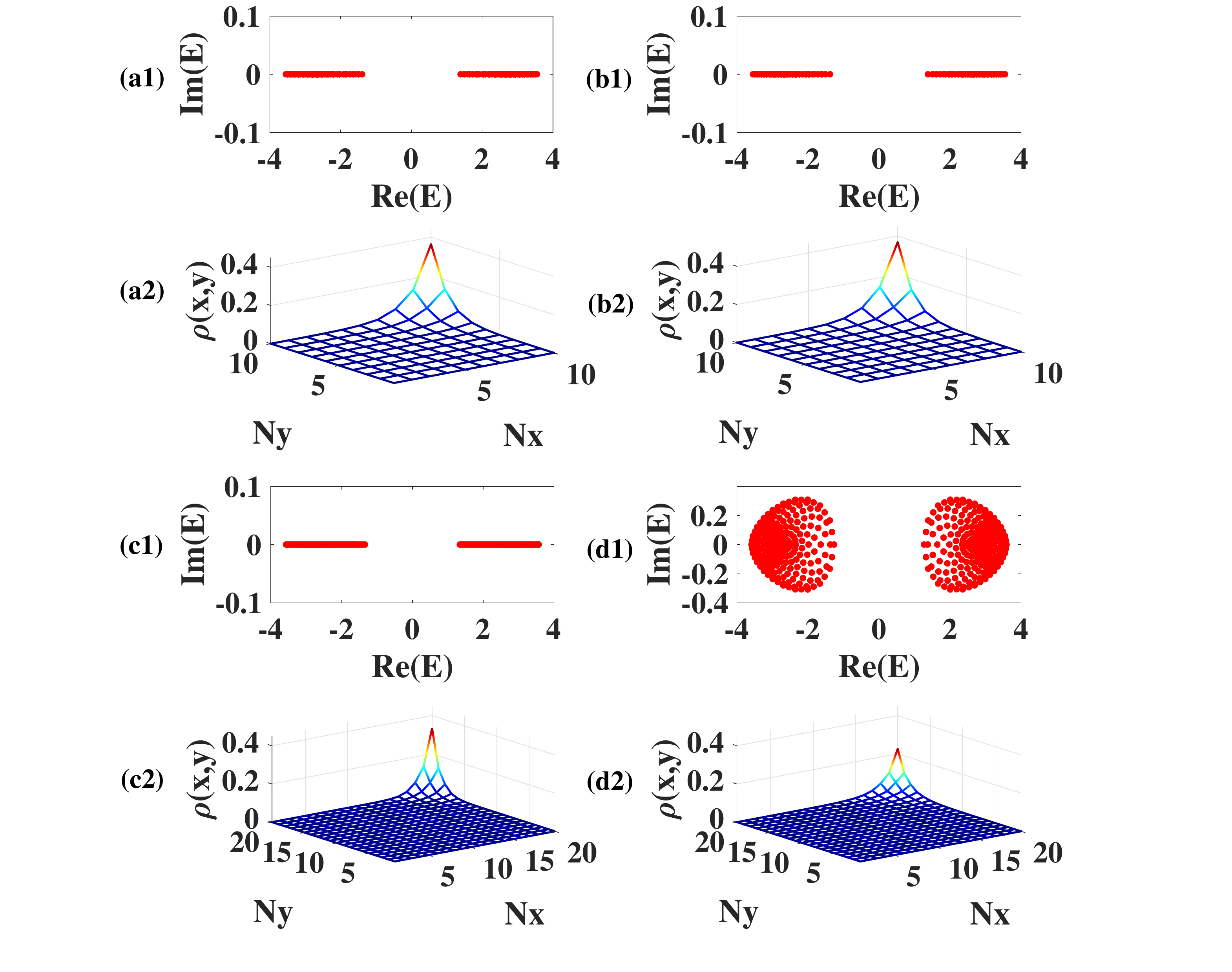}
\caption{
Energy spectra and the profile of eigenstates $\rho(x,y)$  of $\hat{H}_{\mathrm{SOTI}}^{\mathrm{2D}}$ with OBC in both the x-direction and y-direction for systems with size $M_x*M_y=10*10$ (a1,a2) and  $M_x*M_y=20*20$ (c1,c2), respectively. Energy spectra and the profile of eigenstates $\rho(x,y)$  of $\hat{H}_{\mathrm{SOTI}}^{\mathrm{2D}}$ with generalized boundary condition in the x-direction and y-direction ($\delta_{R}^x=\delta_{L}^x=\delta_{R}^y=\delta_{L}^y=0.002$) for systems with size $M_x*M_y=10*10$ (b1,b2) and  $M_x*M_y=20*20$ (d1,d2), respectively.
Common parameters: $t=0.8, \gamma=-1, \lambda=0.8$.}
\label{SOTI2}
\end{figure*}

When we apply PBC in both x-direction and y-direction, the Hamiltonian after Fourier transformation becomes $\hat{H}_{\mathrm{SOTI}}^{\mathrm{2D}}=\sum_{\mathbf{k}}\psi_{\mathbf{k}}^{\dag}H_{\mathrm{SOTI}}^{\mathrm{2D}}(\mathbf{k})\psi_{\mathbf{k}}$ with $\psi_{\mathbf{k}}=(\hat{c}_{\mathbf{k},A},\hat{c}_{\mathbf{k},B},\hat{c}_{\mathbf{k},C},\hat{c}_{\mathbf{k},D})^T$. Here
\begin{equation}
\begin{split}
H_{\mathrm{SOTI}}^{\mathrm{2D}}(\mathbf{k})=[t+\lambda\cos(k_x)]\tau_x-[\lambda\sin(k_x)+i\gamma]\tau_y\sigma_z
+[t+\lambda\cos(k_y)]\tau_y\sigma_y+[\lambda\sin(k_y)+i\gamma]\tau_y\sigma_x,
\end{split}
\end{equation}
where $\tau_i$ and $\sigma_i (i\in x,y,z)$ are Pauli matrices for the degrees of freedom within a unit cell.
When we apply OBC in the x-direction and PBC in the y-direction, the non-Hermitian system supports gapped complex edge states for $|t|<|\gamma|+|\lambda|$, while there are no edge states for $|t|>|\gamma|+|\lambda|$.
When we apply OBC in both the x-direction and y-direction, the non-Hermitian system supports zero-energy corner states in the region of second-order topological phase, and the phase boundaries are determined by $t^2=\lambda^2+\gamma^2$ and $t^2=\gamma^2-\lambda^2$ as demonstrated in Ref.\cite{TLiuSM}.

In order to characterize the amplitude of non-Hermitian skin effect, we defined the averaged squared eigenmode amplitude as
\begin{equation}
\rho(x,y)=\frac{1}{4M_xM_y}\sum_{s}[|\langle x,y,A|\Psi^{s}\rangle|^2+\langle x,y,B|\Psi^{s}\rangle|^2+\langle x,y,C|\Psi^{s}\rangle|^2+\langle x,y,D|\Psi^{s}\rangle|^2],
\end{equation}
where $|\Psi^{s}\rangle$ is the s-th eigenfunctions $|\Psi\rangle$ of $\hat{H}_{\mathrm{SOTI}}^{\mathrm{2D}}$. For simplicity, we focus on the regions in which no gapped edge states or corner states exist, and thus the summation in equation above runs over all eigenfunctions.

When we apply OBC in the x-direction and PBC in the y-direction, the system exhibits non-Hermitian skin effect along x-direction as all wave functions accumulate on the edge, as plotted in Fig.\ref{SOTI1}(a). However, the NHSE along x-direction under tiny boundary perturbations along x-direction is fragile in the thermodynamic limit. As shown in Fig.\ref{SOTI1}(b), the non-Hermitian skin effect along x-direction under tiny boundary perturbations is diminished in comparison with Fig.\ref{SOTI1}(a).

Next, we explore the case with generalized boundary condition in both the x-direction and y-direction numerically.
It is shown that the energy spectra under OBC are always real and there exhibits non-Hermitian skin effect along x-direction and y-direction as all wave functions accumulate on the corner\cite{TLiuSM}, as plotted in Figs.\ref{SOTI2}(a1,a2,c1,c2).
The energy spectra and profiles of eigenstates $\rho(x,y)$ under tiny boundary perturbations is similar to that under OBC for small lattice size as plotted in Figs.\ref{SOTI2}(b1, b2).
As the lattice size $M_x \times M_y$ increases, the non-Hermitian skin effect is diminished under tiny boundary perturbations accompanied by complex spectra as shown in Figs.\ref{SOTI2}(d1, d2). Both the spectra and profiles of eigensates indicate that the non-Hermitian skin effect under tiny boundary perturbations is fragile in the thermodynamic limit for the 2D SOTI. Similar calculation and analysis can be generalized to three-dimensional systems with non-Hermitian skin effect in a straightforward way.


\begin{thebibliography}{99}
\bibitem{note-FGBZ} Finite-size generalized Brillouin zone is introduced to express precisely generalized Brillouin zone for a finite non-Hermitian system.  When $\delta_R=\delta_L=0$ (OBC) and at the large size limit $N \rightarrow \infty$, our solutions $z_1$ and $z_2$ give the generalized Brilouin zone as discussed in Refs.[9,14,15]. Since our solutions hold true for the generalized boundary condition and finite-size systems, we use the term  "finite-size  generalized Brilouin zone", in contrast to the generalized Brilouin zone, which is generally used for non-Hermtian systems under OBC and at the large size limit.
\bibitem{note} Here we keep $\delta_{L,R}$ fixed, i.e., as the values are given, they do not change with the size $N$. So $\eta_2$ always increases exponentially with $N$. However, if we allow the parameters $\delta_{L,R}$ change with $N$, for example, $\delta_{L}/t_L = r^N$ and  $\delta_{R}/t_R = r^{-N}$, then $\eta_2=2$ is a constant and does not change with the increase of N. 
    It is also identical to the case of Eq.(\ref{q1cdt}) with $\mu=r^N$. In this case, we have $\eta_2=1+\eta_1$ ($\eta_1=1$), the solution of $\theta$ can be obtained as $\theta=\frac{2m\pi}{N}, (m=1,2,...,N)$.
\end{thebibliography}
\end{document}